\documentclass[12pt]{article}
\usepackage[pdftex]{graphicx} %%GRAPHICX
\usepackage{epsfig}
\usepackage{comment}
\usepackage{latexsym}
\usepackage{hyperref}
\usepackage{amsmath}
\usepackage{color}

\newcommand{\mysquare}[0]{\raise-.2ex\hbox{{\Large$\Box$}}}
\def\lsim{\mathrel{\rlap {\raise.5ex\hbox{$ < $}}
{\lower.5ex\hbox{$\sim$}}}}
\def\gsim{\mathrel{\rlap {\raise.5ex\hbox{$ > $}}
{\lower.5ex\hbox{$\sim$}}}} \topmargin -1.5cm \textheight=22.5cm \textwidth=16.5cm
\catcode`\@=11
\newcount\hour
\newcount\minute
\newtoks\amorpm
\hour=\time\divide\hour by60 \minute=\time{\multiply\hour by60
\global\advance\minute by-\hour}
\edef\standardtime{{\ifnum\hour<12 \global\amorpm={am}%
        \else\global\amorpm={pm}\advance\hour by-12 \fi
        \ifnum\hour=0 \hour=12 \fi
        \number\hour:\ifnum\minute<10 0\fi\number\minute\the\amorpm}}
\edef\militarytime{\number\hour:\ifnum\minute<10 0\fi\number\minute}
\def\draftlabel#1{{\@bsphack\if@filesw {\let\thepage\relax
   \xdef\@gtempa{\write\@auxout{\string
      \newlabel{#1}{{\@currentlabel}{\thepage}}}}}\@gtempa
   \if@nobreak \ifvmode\nobreak\fi\fi\fi\@esphack}
        \gdef\@eqnlabel{#1}}
\def\@eqnlabel{}
\def\@vacuum{}
\def\draftmarginnote#1{\marginpar{\raggedright\scriptsize\tt#1}}
\def\draft{\oddsidemargin -.2truein
        \def\@oddfoot{\sl preliminary draft \hfil
        \rm\thepage\hfil\sl\today\quad\militarytime}
        \let\@evenfoot\@oddfoot \overfullrule 3pt
        \let\label=\draftlabel
        \let\marginnote=\draftmarginnote
   \def\@eqnnum{(\theequation)\rlap{\k

 ern\marginparsep\tt\@eqnlabel}%
\global\let\@eqnlabel\@vacuum}  }
\newcommand{\be}[0]{\begin{equation}}
\newcommand{\ee}[0]{\end{equation}}
\newcommand{\ba}[0]{\begin{eqnarray}}
\newcommand{\ea}[0]{\end{eqnarray}}

\def\bs{\begin{subequations}}
\def\es{\end{subequations}}

\def\np#1#2#3{Nucl. Phys. {\bf{B#1}} (#2) #3}
\def\pl#1#2#3{Phys. Lett. {\bf{B#1}} (#2) #3}

\def\thebibliography#1{%
\vskip 0.5cm \centerline{\bf \Large References}
\list{%
[\arabic{enumi}]}{\settowidth\labelwidth{[#1]} \leftmargin\labelwidth
\advance\leftmargin\labelsep
\usecounter{enumi}}
\def\newblock{\hskip .11em plus .33em minus .07em}
\sloppy\clubpenalty4000\widowpenalty4000 \sfcode`\.=1000\relax}

\renewcommand{\theequation}{\arabic{section}.\arabic{equation}}

\renewcommand{\section}{\setcounter{equation}{0}\@startsection
{section}{1}{0mm}{-\baselineskip}{0.5\baselineskip} {\normalfont\Large\bfseries}}

\renewcommand{\subsection}{\@startsection
{subsection}{2}{0mm}{-\baselineskip}{0.5\baselineskip} {\normalfont\large\bfseries}}

\renewcommand{\subsubsection}{\@startsection
{subsubsection}{3}{0mm}{-\baselineskip}{0.5\baselineskip}
{\normalfont\normalsize\slshape}}

\usepackage{amssymb}
\usepackage{amsthm}
\usepackage{amsmath}
\usepackage{amssymb,amsfonts}
\usepackage{graphicx}
\usepackage{cite}

\newcommand{\Z}{\mathbb{Z}}

\newcommand{\abs}{|}

\newcommand{\with}{\mbox{with}}

\renewcommand{\and}{\mbox{and}}

\newcommand{\N}{{\cal N}}

\begin{document}
%\verb|\usepackage{draftcopy}|\\
\begin{titlepage}
\begin{flushright}
LPTENS--14/12, LMU-ASC 59/14, MPP-2014-349,
September 2014
\vspace{-.0cm}
\end{flushright}
\vspace{1mm}
\begin{centering}
{\bf \Large ${\cal R}^2$ inflation from scale invariant supergravity}

{\bf \Large and anomaly free superstrings with fluxes }

\vspace{8mm}

 {\bf Costas Kounnas$^{1}$, Dieter L\"ust$^{2,3}$ and Nicolaos Toumbas$^4$}

\vspace{5mm}

$^1$ Laboratoire de Physique Th\'eorique,
Ecole Normale Sup\'erieure,$^\dag$ \\
24 rue Lhomond, F--75231 Paris cedex 05, France\\
{\em  Costas.Kounnas@lpt.ens.fr}

$^2$  {Max-Planck-Institut f\"ur Physik (Werner-Heisenberg-Institut), \\ 
   F\"ohringer Ring 6,  80805 M\"unchen, Germany } \\
   {\em luest@mppmu.mpg.de}
   
%\vspace{0.25cm} 
\emph{$^{3}$ Arnold Sommerfeld Center for Theoretical Physics,\\ 
               LMU, Theresienstr.~37, 80333 M\"unchen, Germany}\\   
{\em dieter.luest@lmu.de}

$^4$  Department of Physics, University of Cyprus,\\
Nicosia 1678, Cyprus.\\
{\em nick@ucy.ac.cy}

\end{centering}
\vspace{0.1cm}
$~$\\
\centerline{\bf\Large Abstract}\\
\vspace{-0.2cm}
%\begin{quote}

\noindent 
The ${\cal R}^2$ scale invariant gravity theory coupled to conformally invariant matter is  investigated. 
We show that in the non-supersymmetric case the conformally coupled scalars belong to an $SO(1, 1+n)/SO(1+n)$ manifold, while in the supersymmetric case the scalar manifold becomes isomorphic to the  K\" ahlerian space ${\cal M}_n$=$SU(1, 1+n)/ U(1)\times SU(1+n)$. In both cases when the underlying scale symmetry is preserved the vacuum corresponds to de Sitter space. 
Once the scale symmetry is broken by quantum effects, a transition to flat space becomes possible. We argue that the scale violating terms are induced by anomalies related 
to a $U(1)_R$ symmetry. The anomaly is resolved via the gauging
 of a Peccei-Quinn axion shift symmetry. The theory describes an inflationary transition from de Sitter to flat Minkowski space, very similar to the Starobinsky inflationary model. 
 The extension to metastable de Sitter superstring vacua is also investigated. The scalar manifold is extended to a much richer manifold, but it contains always ${\cal M}_n$ as a sub-manifold. In superstrings the metastability is induced 
by axions that cure the anomalies in chiral $N=1$ (or even $N=0$) supersymmetric vacua via a Green-Schwarz/Peccei-Quinn mechanism generalized to four dimensions. We present some typical superstring models and discuss the possible stabilization of the no-scale modulus.

%\end{quote}

\vspace{3pt} \vfill \hrule width 6.7cm \vskip.1mm{\small \small \small
  \noindent
   $^\dag$\ Unit{\'e} mixte  du CNRS et de l'Ecole Normale Sup{\'e}rieure associ\'ee \`a
l'Universit\'e Pierre et Marie Curie (Parited s 6), UMR 8549.}\\

\end{titlepage}
\newpage
\setcounter{footnote}{0}
\renewcommand{\thefootnote}{\arabic{footnote}}
 \setlength{\baselineskip}{.7cm} \setlength{\parskip}{.2cm}

\setcounter{section}{0}

\section{Introduction}
It is well known that gravity theories based on higher derivative terms of the type ${\cal R}^n$, where  ${\cal R}$ is the Ricci scalar, are equivalent to standard gravity theories with one additional scalar degree of freedom $\phi$ \cite{Stelle} having a special potential \cite{Starobinsky}. The precise structure of this potential is interesting, since it gives rise to a realization of the inflationary scenario in the early Universe \cite{Guth,REVinflation}. Starobinsky considers the simplest such possibility, adding an additional ${\cal R}^2$ term to the standard Einstein action.  In the conformally equivalent theory to this ${\cal R}+{\cal R}^2$ model, the scalar field $\phi$ has the following potential:
\be
V=  \mu^2\left( 1-e^{-\alpha \phi}\right)^2 \, .
\ee  
The structure of this potential is such that it successfully describes a slow roll transition from a de Sitter phase to a flat Minkowski phase. In fact,
this potential is indeed very suggestive. Firstly, it is semi-positive definite; for large positive values of $\phi$, the potential asymptotes to a vacuum with a positive cosmological constant, while for large negative values the potential grows exponentially. Secondly the global vacuum at $\phi=0$ is stable with zero cosmological term.         

The Starobinsky model can be extended to a phenomenologically viable case by coupling it to matter. Namely, one can couple fermions, gauge bosons and scalar fields to gravity, as required in the realistic world. The precise couplings will play an essential role during the subsequent reheating phase of inflation, and also in the determination of the ratio $r$ of the tensor versus the scalar perturbations \cite{REVinflation}. In the Starobinsky model, the scalar perturbation fixes the scale of inflation to $\mu\sim 10^{13}\, GeV$. Furthermore it apparently gives rise to a relatively small ratio $r\sim 10^{-3}$. These results are not in contradiction with observations by the Planck and other astrophysical experiments \cite{PLANCK}, where they provided some upper bounds on the amplitude of the tensor perturbations \cite{REVinflation}. However, recent data from the BICEP2 astrophysical experiment suggest a significantly bigger value, namely $r\sim 0.2$ \cite{BICEP2}. If this relatively large value turns out to be further confirmed by other astrophysical observations, the minimal Starobinsky model will be disfavored. Hence a suitable generalization to an inflationary model with a richer structure will be necessary.

Several works in the literature focus on the supergravity embedding of the Starobinsky model 
\cite{Cecotti,N=1susyStarobinsky,DtermsOnly,Ellis,Lahanas:2014ula,Antoniadis,FerraraVanProyen,NewFerraraPorrati,N=2susyStarobinsky}. These supergravity extensions not only concern the $N=1$ super-symmetrization, but also the super-symmetrization at the $N=2$ level of the theory \cite{N=2susyStarobinsky} (to our knowledge the extension for  $N>2$ has not yet been investigated), where all these extensions are performed at the classical level. 
In a sense, they are still 
incomplete   for the description of
a realistic world, since $N=1$ supersymmetry has to be spontaneously broken to $N=0$, including also
 the spectrum of the  Standard Model of particle interactions.  

The scope of our work is to understand the origin of the approximate de Sitter phase of the ${\cal R}+{\cal R}^2$ theory based on symmetry principles. We begin with the observation that the pure 
${\cal R}^2$ theory describes an exact de Sitter space as a consequence of an unbroken global scale symmetry. Once this symmetry is softly broken by the standard Einstein term and/or by other mass terms, then a slow ``inflationary" transition towards flat Minkowski space becomes possible. In order to understand the (quantum) origin of various scale violating terms, we generalize, as a first step, the pure ${\cal R}^2$ theory by adding extra matter fields, like conformally coupled scalars, fermions and gauge bosons, obtaining in this way a scale invariant theory with a non trivial interacting structure, very similar to that of the low energy Standard Model. It is then clear, even at this step, that the scale violating terms, which are forbidden at the classical level thanks to scale invariance,  will be induced at the quantum level of the theory, where conformal and scale invariance are broken in a more or less controllable way (at least for the matter sector of the theory). As a second step, we consider the $N=1$ supergravity extension of the ${\cal R}^2$ theory. This will give us further insight into the structure of the scale violating terms. 
In particular the undressed  Einstein term ${\cal R}$, which violates scale invariance, 
is induced at the quantum level of the theory via an anomaly cancellation condition involving
 a $U(1)_R$ gauge symmetry. The corresponding Fayet-Iliopoulos $D$-term \cite{fayet} gives rise to 
the de Sitter cosmological contant \cite{StelleFI,FabioDeSitter}. 
We argue that the anomaly is resolved by gauging simultaneously
a Peccei-Quinn axion shift symmetry, or via the 
Green-Schwarz mechanism \cite{GSAnomalyBmn} adapted to four dimensions 
\cite{FIstrings,FIbranesIbanez,ChiralAndU(1)R}, similar to the global supersymmetric case\cite{GlobalSusy}.\footnote{The special role of anomalous $U(1)$'s 
in moduli stabilization and inflation was also discussed in \cite{li}.}

Over the last years the derivation of de Sitter vacua and the embedding of inflationary effective supergravity potentials to superstrings 
was discussed in many papers (see e.g.    \cite{Kachru:2003aw,Kachru:2003sx}  and for recent reviews on this subject   
 \cite{Burgess:2013sla,Linde:2014nna,Baumann:2014nda,Kallosh:2014xwa}).
For orientifolds with fluxes and branes even some no-go theorems against stable de Sitter vacua were formulated \cite{Hertzberg:2007wc}
and ways to bypass these arguments were subsequently discussed in 
\cite{Caviezel:2008tf,Danielsson:2009ff,Danielsson:2009ff,Haque:2008jz,Danielsson:2011au,Blaback:2013ht,Damian:2013dq,Damian:2013dwa,Hassler:2014mla}.
In our paper we will show how the scale symmetry of the ${\cal R}^2$ supergravity effective actions including 
matter as well as the non-scale invariant gaugings from
quantum effects can be naturally derived from string compactifications
with fluxes. In this way we will describe a new way to obtain de Sitter vacua and inflation from underlying superstring constructions.
In fact, the structure of superstring effective supergravities (including matter) is by now well understood 
(at least in orbifold and orientifold compactifications with fluxes), not only when $N=1$ supersymmetry is preserved, 
but also when $N=1$ supersymmetry is spontaneously broken to $N=0$ via geometrical fluxes. In many superstring compactifications the degrees of freedom 
are organized in sectors with higher non-aligned supersymmetries.      
This  is also valid in  $N=1\to N=0$ four-dimensional string compactifications. This is the situation relevant for our discussion and it is realized in many
four-dimensional 
%$Z^2\times Z^2$ orbifold compactifications of the 
heterotic superstring constructions \cite{fermionic,Lerche:1986cx,orbifolds,AsOrbifolds}, where the six-internal coordinates define three ``twisted'' complex planes, 
and where the states (and the quantum corrections associated to them) are organized in sectors with higher supersymmetries: one $N=4$ sector and three non-aligned $N=2$ supersymmetric sectors. 
%The $N=4$ sector is the remnant of fields of the initial $N=4$ theory surviving the orbifold projections, while in the $N=2$ sectors there are extra twisted states 
%\cite{fermionic,orbifolds,AsOrbifolds}
%which give rise to the $N=1$ chiral states. 
Furthermore, the situation often occurs in orientifold constructions of the type II superstrings, where now the chiral states are localized on $D$-branes 
(for a review on orientifolds with $D$-brane see e.g. \cite{Blumenhagen:2006ci}). 
In supergravity theories, such orbifold constructions result in anomalous theories. The string effective supergravities however are anomaly free thanks to the additional ``twisted'' \cite{FIstrings,ChiralAndU(1)R,N=0Sectors} or localized $D$-brane states \cite{FIbranesIbanez}. 
Therefore, the study of the vacuum structure of
an $N=1\to N=0$ chiral model \cite{N=0Sectors}, as well as the study of the underlying $N=4$ and $N=2$ sectors, 
are necessary for obtaining quantitative control over the 
quantum corrections of the theory. 
%For applications to cosmology, this study can prove to be very useful, 
%if in particular the inflaton field happens to belong to a sub-sector of the underlying 
%fundamental string theory with $N>1$ supersymmetry.

The paper is organized as follows. In the next section we first briefly recall the structure of Einstein gravity plus an additional ${\cal R}^2$ interaction, and also the minimal version of a scale invariant ${\cal R}^2$ theory 
without Einstein term. This theory will then be extended by introducing conformally coupled scalar fields, which is equivalent in the Einstein frame to standard gravity
with a positive cosmological constant plus scalar fields with a quartic potential with constant scalar mass terms. After breaking the scale invariance, a dynamical transition from the scale invariant de Sitter phase to the
non scale invariant flat phase occurs via a dynamical inflationary potential. 
In sections three and four the $N=1$ supergravity extension of these theories will be discussed in some detail, 
where the matter part of
the supergravity action in general is built from simultaneous $F$- and $D$-terms.
Here the form of the possible $U(1)$ $R$-symmetry gaugings plays a crucial role for the determination of the final scalar potential
and its dynamical stability properties, as it can be done either in classical, however anomalous fashion (called $U(1)_R$ gauging), but also in two
non-anomalous ways, which are related to the quantum contributions
of the theory. Even more interestingly, these two non-anomalous gaugings can be performed scale invariantly (called $U(1)_d$ dilatational
gauging) but also in a scale violating way (called $U(1)_t$ translational
gauging), which reflects the structure of the non-scale invariant theories of section 2.2, and which leads to inflationary scalar potentials. Finally in section 5, the string derivation of
these supergravity theories from compactifications with fluxes 
will be discussed.

\section{The ${\cal R}^2$ model plus conformally coupled matter}

\subsection{Preliminaries}
It is well known that gravity theories with higher derivative terms of the type ${\cal R}^n$, where  ${\cal R}$ is the Ricci scalar, are equivalent to standard gravity theories with one extra scalar degree of freedom $\phi$.
Starobinsky considers the simplest such possibility \cite{Starobinsky}, adding an additional ${\cal R}^2$ term to the Einstein action\footnote{ We are working in the reduced Planck units with $M=2.4 \times 10^{18}\, GeV$.}:
\be
S=\int d^4x\sqrt{\abs g \abs} {1\over 2}\left( {\cal R} +{{\cal R}^2\over 8 \mu^2} \right)\, .
\ee
Introducing a Lagrange multiplier $t$, one can replace the $R^2$ term with,  
$$
2t{\cal R}-8\mu^2 t^2 \, , 
%~\longrightarrow~ t={R \over8 \mu^2}\, ,
$$
which after the $t$ equation of motion, $t={{\cal R} /8 \mu^2}$, reproduces the initial ${\cal R}^2$ term. Thus the action can be written in an equivalent way in terms of the scalar field $t$:
\be
S=\int d^4x\sqrt{\abs g \abs} {1\over 2}\left\{(1+2t) {\cal R} -8\mu^2 t^2   \right\}\,. 
\ee
In this Jordan frame, $t$ looks like a non-propagating field. This however is an illusion of this frame. Performing a rescaling of the metric in order to obtain a canonical Einstein term,
$$
g_{\mu \nu} \rightarrow g_{\mu \nu}\,e^{-\log(1+2t)}\, ,
$$
the action takes the form:
\be
S=\int d^4x\sqrt{\abs g \abs} ~\left[{1\over 2} {\cal R} -3{ \partial_{\mu} t \partial^{\mu} t\over (1+2t)^2}   
-\mu^2 { \left( 2t\right)^2  \over (1+2t)^2 } \right]\, ,
\ee
\be
S=\int d^4x\sqrt{\abs g \abs} ~{1\over 2}\left[ {\cal R} - \partial_{\mu} \phi \partial^{\mu}\phi -   
2\mu^2\left( 1-e^{-\alpha \phi}\right)^2   \right]\, ,
\ee
where in the second equation above we introduced the canonically normalized scalar field $\phi$,
\be
\alpha \phi=\log(1+2t)  ~~{\with}~~~ \alpha= \sqrt{2\over 3}~ .
\ee
To obtain the above expressions, we used the following rescaling formula in d-dimensions:
$$
\int d^dx\sqrt{\abs g \abs} ~{1\over 2}Y {\cal R} \rightarrow \int d^dx\sqrt{\abs g \abs} ~\left({1\over 2} {\cal R} -{(d-1)\over 2(d-2)}\,  {\partial_{\mu}Y \partial^{\mu}Y\over Y^{d\over d-2}}\right)\, .
$$
Therefore in the dual expression of the ${\cal R}+{\cal R}^2$ theory (when written in the Einstein frame), the canonically normalized scalar field $\phi$ admits a very special potential:  
\be
V=  \mu^2\left( 1-e^{-\alpha \phi}\right)^2 \, .
\ee
Notice that the potential is semi-positive definite as soon as $\mu^2>0$. 
For large positive values of $\phi$, $V$ asymptotes to a constant, while for large negative values the potential grows exponentially. The global minimum is at $\phi=0$, where the potential vanishes. In summary
\be
V(\phi\gg 0)
\rightarrow  {\mu^2},~~~~~~V(\phi= 0)=0, ~~~~~V(\phi\ll0)\sim {\mu^2}e^{-2\alpha\phi}\, .
 \ee

Hence for large positive values of $\phi$, the vacuum is described by an approximate de Sitter space, which grows exponentially with Hubble parameter proportional to $\mu$,
\be
3H^2={\mu^2}\,,~~~~~~~~~~~ H={\dot a\over a}\, .
\ee
Here $a$ denotes the scale factor of the metric $ds^2=-dt^2+ a^2(dx_i)^2$. For this reason, the ${\cal R}+{\cal R}^2$ theory was proposed by Starobinsky to describe inflationary cosmology. In order to create a successful density perturbation, 
${\delta \rho / \rho} \sim 10^{-5}$, the mass scale $\mu$ must be as low as $\mu\sim 10^{-5}\, M\sim 10^{13} GeV $.

Before coupling the model to realistic matter, we consider the pure ${\cal R}^2$ theory, which represents a {\it minimal version of a scale invariant theory without ghosts}:  
\be
S=\int d^4x\sqrt{\abs g \abs} {1\over 2}\left( {{\cal R}^2\over 8\mu^2} \right)\,~~~~\longrightarrow~~~~
S=\int d^4x\sqrt{\abs g \abs} {1\over 2}\left[2t{\cal R} -8\mu^2 t^2   \right]\,.
\ee
The Lagrange multiplier $t$ is introduced as previously, in order to re-express the theory in the standard form linear in ${\cal R}$. 
The important observation is that the ${\cal R}^2$ theory is invariant under global dilatations:
\be
g_{\mu \nu} \rightarrow e^{-2\sigma} g_{\mu \nu}, ~~~~t \rightarrow e^{2 \sigma} t \, ,
\ee
where $\sigma$ is constant. Performing the conformal rescaling of the metric, in order to pass from the Jordan to the Einstein frame, and performing a field
redefinition 
  $\alpha \phi=  \log(2t)$ with $ \alpha= \sqrt{2\over 3}$,
we obtain
\be
S_E=\int d^4x\sqrt{\abs g \abs} ~{1\over 2}\left[ {\cal R}  - \partial_{\mu} \phi \partial^{\mu}\phi - 2\mu^2 \right]\, .
\ee
We observe that the pure ${\cal R}^2$ theory is conformally equivalent to an Einstein gravity theory with a positive cosmological constant,  coupled to a massless scalar field $\phi$. The vacuum solution to this theory is nothing but de Sitter space.  In the Einstein frame,  the initial dilatation symmetry is translated to a shift symmetry acting on $\phi$, $\alpha\phi\rightarrow \alpha\phi +2\sigma$, while the metric remains invariant. Because of this symmetry the only possible potential for $\phi$ is a cosmological constant $\mu^2$, which can be positive (de Sitter), negative (Anti de Sitter) or in the limiting case zero (flat space).

\subsection{ The $SO(1,1+n)$ generalization of the pure ${\cal R}^2$ theory} 
We generalize the pure ${\cal R}^2$ theory by adding extra matter fields like conformally coupled scalars, fermions and gauge bosons so as obtain a more realistic scale invariant theory with a non trivial interacting structure\footnote{The ${\cal R}^2$  theory of gravity coupled to matter was previously also considered in
 \cite{Salvio:2014soa}.}
It is well known that attributing to the conformally coupled scalars a scaling weight $w_s=1$, to the fermions $w_f=3/2$ and to the gauge bosons $w_g=0$ then, the global scale symmetry of the matter sector is promoted (at the classical level of the theory) to a local conformal symmetry, provided that the interactions are gauge or Yukawa type and that the scalar potential is quartic, $V_c=\lambda_{ijkl}~\Phi_i\Phi_j\Phi_k  \Phi_l$. 

At the classical level, the scaling symmetry forbids any kind of mass terms and trilinear scalar interactions. It is well known however that such scale violating terms will be induced at the quantum level of the theory since then scale invariance is broken. Nowadays however, the quantum scale violating effects are more or less well understood. They give rise to non trivial coupling constant renormalization for the interactions and non trivial anomalous dimensions for the fields and other operators. Also, they are strongly restricted by all possible anomalies that can appear in local interactions. It makes sense therefore to start with a theory which is scale invariant at the classical level and then implement the quantum corrections \cite{CollemanWeiberg}, especially the requirements related to anomalies as a second step. 

With this in mind, the first minimal extension of the ${\cal R}^2$ theory is achieved by introducing {\it conformally coupled  matter}. The fermionic and gauge boson parts of the action are invariant under conformal transformations. The action is
\be
S=\int d^4x\sqrt{\abs g \abs} {1\over 2}\left( {{\cal R}^2\over 8 \mu^2} -{1\over 6}\Phi_i^2 {\cal R}- \partial_{\mu} \Phi_i \partial^{\mu}\Phi_i-2V_c(\Phi_i)+\dots \right)\, ,
\ee
where the ellipses denote the fermionic and gauge boson matter parts. These parts do not affect the vacuum structure of the model. Observe the absence of the canonical Einstein term ${\cal R}$, since it is not permitted by the scale symmetry. The linear ${\cal R}$ term is dressed by the fields $\Phi_i$ which have conformal weight 
$w_{\Phi}= 1$.
%$\Phi_i^2R$.
Introducing as previously the Lagrange multiplier field $t$, the action in Jordan frame takes a more familiar form, linear in ${\cal R}$:
\be
S=\int d^4x\sqrt{\abs g \abs} {1\over 2}\left(\left[2t-{1\over 6}\Phi_i^2 \right]{\cal R}- \partial_{\mu} \Phi_i \partial^{\mu}\Phi_i-2V_c(\Phi_i)-8\mu^2 t^2   \right)\, .
\ee
 Defining
 \be \label{scalarphi}
  \alpha \phi=  \log\left(2t-{1\over 6}\Phi_i^2\right)  ~~{\with}~~~ \alpha= \sqrt{2\over 3}\, ,
 \ee
 and performing a conformal rescaling of the metric 
 $$
   g_{\mu \nu} \rightarrow  g_{\mu \nu}\,e^{-\log(2t-{1\over 6}\Phi_i^2 )}\, ,
 $$
we obtain the action in the Einstein frame:
\be
S_E=\int d^4x\sqrt{\abs g \abs} ~{1\over 2}\left[ {\cal R} - \partial_{\mu} \phi \partial^{\mu}\phi  - e^{-\alpha \phi}\partial_{\mu} \Phi _i\partial^{\mu}\Phi_i  
-2e^{-2\alpha \phi }V_c(\Phi_i)-2\mu^2\left( 1+{e^{-\alpha \phi}\Phi_i^2\over 6}\right)^2   \right]\, .
\ee

Some comments are in order:
\begin{itemize}
\item In the Einstein frame the metric is invariant under scale transformations due to the conformal dressing.
\item The full classical action is invariant under the following scale symmetry:
\be
\alpha \phi\rightarrow \alpha \phi+2\sigma \, ,~~~~~~\Phi_i\rightarrow e^{\sigma}\Phi_i\, ,~~~~~
g_{\mu \nu}\rightarrow g_{\mu \nu} \, ,
\ee 
provided the potential $V_c$ is quartic: $V_c=\lambda_{ijkl}~\Phi_i\Phi_j\Phi_k \Phi_l$. The $\mu^2$ term induced by the initial ${\cal R}^2$ theory is also invariant. It contains a constant term giving rise to a positive cosmological constant plus $\Phi^2$ mass and $\Phi^4$ interaction terms, dressed by powers of $e^{-\alpha \phi}$ in a way that renders them scale invariant. 
\item The above scale symmetry is extended to the fermionic and gauge interacting parts of the theory as a remnant of the initial conformal invariance of the matter sector.
\item The scalar kinetic terms indicate that the scalar manifold ${\rm\cal M} (\phi, \Phi_i)$, $i=1,...n$ is isomorphic to a maximally symmetric space, namely the hyperbolic space ${\rm \cal H}^{n+1}$ (the Euclidean $AdS$ space):
\be
{\rm\cal M} (\phi, \Phi_i)={\rm\cal H}^{n+1}\equiv {SO(1,1+n)\over SO(1+n) }.
\ee
\item As in the case of the pure ${\cal R}^2$ theory, here also, the solution to the classical equations of motion is de Sitter space, where $\phi$ must be constant and $\Phi_i=0$. The cosmological constant is positive and equal to $\mu^2$. 
\item The induced mass terms for the scalars $\Phi_i$, are of the form expected for conformally coupled scalars in a de Sitter background: $m^2_i={1\over 6}<{\cal R}>$. Indeed in the Jordan frame the effective (mass)$^2$ terms of the scalars are 
proportional to $<{\cal R}>$. Translating into the Einstein frame and taking into account the normalization of the kinetic terms of the fields, gives $m^2_i={1\over 6}<{\cal R}_E>\abs_{\Phi_i=0}={2\over 3}\mu^2$. These are precisely the mass terms derived from the induced $\mu^2$ part of the potential in the Einstein frame.
\end{itemize}

\subsection{The $SO(1,1+n)$ model with a  scale violating term }
\label{soR}
The additional $R$ term in the action breaks the classical scale invariance since it shifts the field $2t\rightarrow 2y=2t+1$. 
More explicitly in terms of the field $\phi$ associated to the conformal rescaling of the metric,
$$
\alpha \phi= \log\left(2t+1-{1\over 6}\Phi_i^2\right)=\log\left(2y-{1\over 6}\Phi_i^2\right)\, ,
$$
the $\mu$-dependent part of the Einstein frame potential gets a contribution from a scale non-invariant term:
\be
V_{\mu^2}={\mu^2}\left( 1+{e^{-\alpha \phi}\Phi_i^2\over 6}\right)^2 ~~~ \longrightarrow ~~~V_{\mu^2}={\mu^2}
\left( 1-e^{-\alpha \phi}+{e^{-\alpha \phi}\Phi_i^2\over 6}\right)^2  \, .
\ee
Notice that the structure of the kinetic part of the action remains unaltered, defining a scalar manifold which is isomorphic to the 
${\cal H}^{1+n}=SO(1,1+n)/SO(1+n)$ Hyperbolic space. Therefore the scale violating term ${\cal R}$ modifies the potential but not the kinetic part of the theory.

 The effect of the scale violating term Einstein term, or equivalently of the term $\{-e^{-\alpha \phi}\}$ in the potential, is very essential, as it gives rise to an inflationary vacuum structure, very similar to the one initially proposed by Starobinsky. It changes drastically the vacuum, inducing a transition from the scale invariant de Sitter phase with $<{\cal R}>=4\mu^2$ to the flat phase with $<{\cal R}>=0$. Notice that in the initial scale invariant theory the explicit dependence of the potential on $\phi$ can be always absorbed into the canonically normalized, scale invariant fields $\hat\Phi_i=e^{-{\alpha \over 2}\phi}\, \Phi_i$. Writing the theory in terms of $\hat\Phi_i$, the field $\phi$ appears only through its spacetime derivatives. Therefore the potential is $\phi$ independent as soon as the theory is scale invariant. As a result the field $\phi$ remains always a flat direction when the scale symmetry is preserved.

 Once scale violating terms are introduced, the potential acquires $\phi$ dependence, and the vacuum structure of the theory gets modified drastically. On the other hand the scale violating terms, like for instance the Einstein term ${\cal R}$ we are considering here, are very welcome since they induce a dynamical transition from the initial de Sitter phase to a more realistic phase, able to describe our Universe today. The fundamental question that we would like to investigate concerns the quantum origin of such scale violating terms, viewed in the framework of a more fundamental theory, like supergravity or superstring theory. Indeed this is a difficult task, which cannot be implemented without extra input from a fundamental theory of quantum gravity. We will not be able to claim anything about a possible quantum origin of the ${\cal R}$ term in the framework of the non-supersymmetric $SO(1,1+n)$ model. We will be able however to go much further, once we consider its supersymmetric extension, which we study in more detail in the following sections. Here we analyze the $SO(1,1+n)$ extension of the ${\cal R} ^2$ theory, in the presence of the ${\cal R}$ scale violating term, without asking for the moment about its quantum origin. This analysis will be useful not only for the $SO(1,1+n)$ model, but also for all its supersymmetric extensions that we consider in the next sections.  

In what follows we study the three possible phases of the $SO(1,1+n)$ model, which are emerging in the presence of the scale violating term ${\cal R}$, equivalently in terms of the expectation values of the fields $\phi$ and $\hat \Phi_i$. 
Generically there are three phases: 

(i)  $~\phi\gg 0$ : when the scale breaking terms are negligible.

(ii) $\phi\sim 0$ : when both the scale breaking terms and the scale invariant terms\\ 
$~~~~~~~~~~$ are of the same order of magnitude.

(ii) $\, \phi\ll 0$ : when the scale breaking terms are dominant.

\subsubsection{(i) The $ SO(1, 1+n)$ scale invariant de Sitter era}
When $\phi> 0$, the induced $({\rm mass})^2$ of the canonically normalized fields $\hat \Phi_i$ in an approximate de Sitter background are positive definite: 
$m_i^2={2\mu^2 \over 3}(1-e^{-\alpha \phi})$.
For large positive values of $\phi$, $V_{\mu^2 }$ is approaching to a constant, $V_{\mu^2} \rightarrow \mu^2$, modulo the massive excitations of the $\hat \Phi_i^2$ fields. These are frozen at $\hat \Phi_i^2=0$ independently of the structure of $ V_c(\hat \Phi_i)$, which is quartic in $\hat \Phi_i$. This statement remains valid even in the presence of {\it mass scale breaking terms}, which are induced at the quantum level of the theory, provided that $(m^i_q)^2+{2\mu^2 \over 3}>0$. Differently stated, the Coleman-Weinberg quantum $(\rm mass)^2$,  $(m^i_q)^2$ (which in general can be negative) \cite{CollemanWeiberg}, has to be smaller (in absolute values) than the effective de Sitter $({\rm mass})^2={\, 2\mu^2/3}$ .  In the case where a particular $\hat \Phi_i$ direction is protected by the quartic terms of $V_c$, the bound on the quantum generated mass terms is not even necessary. The only thing that can happened is that this particular $\hat \Phi_i$ will develop a non-trivial vacuum expectation value during the inflationary era. This can modify the effective value of $\mu^2$, but it can not modify entirely the inflationary behavior of the theory.

\subsubsection{(ii) The $ SO(1, 1+n)$ flat space era}
The potential is semi-positive definite a fact that simplifies the analysis. For any vacuum the total potential vanishes:   
$V=V_{\mu^2}(\phi, \hat \Phi_i)+V_c(\hat \Phi_i)=0$ (zero effective cosmological term). Furthermore, both 
parts of the potential, $V_c$ and $V_{\mu^2}$, must be zero separately. The symmetric point $\hat \Phi_i=0$, 
$\phi=0$ is always a minimum independently of the choice of $V_c$. In the case where $V_c$ does not have flat directions, 
then the symmetric vacuum is unique. When there are flat directions in $V_c$, then there is a degeneracy vacua along the flat direction of $V_{\mu^2}$:
\be
<{\hat \Phi}^2_i= e^{-\alpha \phi}-1>, ~~~~{\rm with}~~~V_c=V_{\mu^{2}}=0 \, .
\ee

\subsubsection{ (iii) The $ SO(1, 1+n)$ scale non-invariant era}
Here also the behavior of the potential depends crucially on the structure of $V_c$, namely whether there are flat directions or not.
Suppose there are no flat directions. Then we can write the potential in terms of the radial field  $\rho^2={\abs \hat \Phi_i \abs^2 \over 6}$:
\be
V=h^2\rho^4 +\mu^2\left\{\rho^2 -(e^{-\alpha \phi}-1)\right\}^2, ~~{\rm with}~~~ (e^{-\alpha \phi}-1)>0\, ,
\ee
where $h^2$ is given in terms of the direction in the field configuration space of $\hat \Phi_i $ (modulo their coupling constants) which gives the lowest contribution to $V_c$. Minimizing further with respect to $\rho$ (while keeping $\phi$ arbitrary with $(e^{-\alpha \phi}-1)>0$), the off shell $\phi$-potential becomes:
\be
V={h^2\mu^2\over h^2+\mu^2} \left(e^{-\alpha \phi}-1\right)^2 \, , ~~~~\rho^2 ={\mu^2\over h^2+\mu^2}\left(e^{-\alpha \phi}-1\right).
\ee
 Therefore, the structure of the potential (in the phase where the breaking terms dominate), turns out to be the same as in the minimal Starobinsky model, even thought in this phase additional matter fields have non-trivial expectation values. Notice that at the extremum $\hat \Phi_i=0$, the potential 
 \be 
V(\hat\Phi_i=0)=\mu^2 \left(e^{-\alpha \phi}-1\right)^2 \, ,
\ee
is always higher than the minimum with $\rho\ne 0$, showing that the extremum at  $\rho^2 =\abs\hat \Phi_i\abs^2/6=0$ is an unstable maximum.
 
When $V_c$ has flat directions, then the effective coupling $h^2$=0.
In this case the total off-shell potential is fully controlled by $V_{\mu^2}$:
\be
V=\mu^2\left\{\rho^2 -(e^{-\alpha \phi}-1)\right\}^2\, .
\ee
Contrary to the previous non degenerate case with $h^2\ne 0$, the vacuum structure differs drastically from that of the Starobinsky model. The explicit presence of $\rho$, which is not stabilized by $V_c$, screens the exponential behavior of $\phi$ along the flat direction $\rho^2 =(e^{-\alpha \phi}-1)$ with vanishing total potential. Here, the effects of additional scale breaking terms emerging at the quantum level of the theory, like for instance Coleman-Weinberg mass terms for the fields $\hat \Phi_i$, will play an important role, since in the effective quantum potential of the theory both $\rho$ and $\phi$ will be fixed \cite{sun1noscale,noscale,FKZ}.  

Before moving to the supersymmetric realization of the $SO(1,1+n)$ model we would like to summarize the main observations concerning the scale breaking effects.
\begin{itemize}
\item In the absence of scale violating terms, the inflationary de Sitter era is stable. The fields $\hat \Phi_i$ receive an effective mass which stabilizes the radial 
direction at $\rho^2=0$.
\item The presence of the scale violating term ${\cal R}$ destabilizes the de Sitter vacuum, and a transition occurs towards flat Minkowski space. 
Rolling to the flat vacuum is slow since the scale violating terms in the initial approximate de Sitter era are exponentially small. 
In this regime, the semiclassical approximation can be trusted.  
\item The quantum origin of the ${\cal R}$ term must be discussed in the context of a more fundamental theory of quantum gravity. 
\item At the end of the inflationary period, the vacuum is approximately flat Minkowski space. At the classical level the vacuum structure can be degenerate 
with flat directions and vanishing cosmological term.
\item At the quantum level the degeneracy will be lifted by the induced scale violating terms in the quantum effective potential. 
These effects are more or less controllable around an approximate Minkowski background especially in supersymmetric theories.
\item To understand the quantum origin of the gravitational scale breaking terms we need to move to more fundamental theories, like for instance superstring theories.  
\end{itemize}

We proceed in the following sections in some interesting 
supersymmetric and superstring generalizations of the $SO(1,1+n)$ ${\cal R}^2$ inflationary model.

\section{Supergravity extensions of the ${\cal R}^2$ theory}
There is extensive work in the literature concerning the supergravity extensions 
of the Starobinsky model. While some of our results have been obtained in the past (see in particular \cite{NewFerraraPorrati}), our approach and set up are new. 
Our main motivation in this work is to construct
scale invariant theories in a de Sitter background, extending the ${\cal R}^2~\oplus$ conformally invariant matter theories in the framework of $N=1$ supergravity, investigating at the same time their connections with gauged supergravity and superstring effective theories with fluxes.

The scale non invariant terms will be systematically added, according to the consistency of the theory at the quantum level. In particular we take into account anomaly cancellation conditions associated with local gauge symmetries and their connection with the induced ${\cal R}$ scale breaking term; quantum induced mass scale violating terms; finally in the presence of spontaneous supersymmetry breaking, the effective soft supersymmetry breaking terms. 

 \subsection{The minimal scale invariant $SU(1,1+n)$ supergravity extension }
 The minimal supersymmetric extension of the non-supersymmetric  $SO(1,1+n)$ model can be realized by introducing the supersymmetric scalar partners of $t$ and $\Phi_i$ appearing in the $SO(1, 1+n)$ ${\cal R}^2$-model. Thus we introduce the complex field $T$ and the fields $ z^i,\, i=1,...n$ so that
 \be 
 T=t+ib ~~~~~z^i=\abs z^i \abs e^{i\theta^i}, ~~~\abs z^i \abs ={\Phi^i \over \sqrt{6}} \, .
 \ee
 In terms of $T$ and the $z^i$'s the supergravity action is given by (here $\psi^I=\{T, z^i\}$)
 \be
 S=\int d^4x\sqrt{\abs g \abs} ~\left[ {1\over 2} Y \left({\cal R}+{2\over 3}A_{\mu}A^{\mu}\right) -J_{\mu}A^{\mu} +3Y_{I\bar J}\,\partial_{\mu}\psi^I \partial^{\mu}{\bar \psi}^{\bar J}- V_c\right] ,
 \ee
 where
 \be
 Y=T+\bar T-\abs z^i\abs^2 \, ,~~~ Y_I={\partial Y\over \partial \psi^I}\, ,~~~Y_{\bar I}={\partial Y\over \partial \bar \psi^{\bar I}}\, ,~~~~Y_{I\bar J}={\partial  Y\over \partial \psi^I \partial \bar \psi^{\bar J}}\, .
 \ee
The supergravity construction we are using in this work is based on the so called ``old minimal formalism for the supergravity multiplet''  as described by Cremmer {\it et al} \cite{MinimalSUCRA}. We will not need to go beyond the standard supergravity description at any point in this work.
The relation between different auxiliary field formulations of $N=1$ supergravity coupled to matter \cite{newSUGRA} can be useful in certain cases, like for instance in solving the apparent non-holomorphic structure of the the gauge kinetic function $f^{ab} (z^I)$, after incorporating the string threshold corrections to the gauge couplings\cite{thresholds}. The new set of auxiliary fields resolves this obstruction in an elegant way \cite{FerraraKounnasDerendingerZwirner}. The ``old minimal formalism'' is much more general than the ``new minimal one''. The two  become equivalent in certain cases by means of a duality transformation 
of the antisymmetric tensor fields to axions in a supersymmetric way \cite{NewFerraraPorrati,FerraraKounnasDerendingerZwirner}.

In the action above the contribution of the auxiliary vector field $A_{\mu}$ appears naturally together with the Einstein term ${\cal R}$. This vector field $A_{\mu}$ is necessary for the supersymmetric extension of the model being a member of the ${\cal R}$ supermultiplet.  Neglecting fermionic contributions, $A_{\mu}$ is determined in terms of the axial current $J_{\mu}$ via its algebraic equation motion \cite{MinimalSUCRA}. In all, $A_{\mu}$ and $J_{\mu}$ are given in terms of the scalar fields by the following expression: 
 \be
 A_{\mu}={3\over2}{J_{\mu}\over Y} ={3\over2}i{Y_{\bar I}\partial_{\mu} { \bar \psi}^{\bar I}-Y_{ I } \partial_{\mu}\psi^{ I}\over Y}, ~~~~~~
 \psi^I=\{T,z^i\}\, .
 \ee
 As in the cases of the pure $R^2$ scale invariant model and its extension to the $SO(1,1+n)$ model, the field $T+\bar T=2t$ appears as a non-propagating field in the Jordan frame. Integrating this out gives rise to an equivalent theory with non trivial ${\cal R}$ dependence. Indeed, the $t$ equation of motion yields
 \be
{\cal R} ={\partial \over \partial t}V_c(T, z^i) \, ~~~ \rightarrow~~~ T+\bar T=f({\cal R}, b, z^i)\, ,
 \ee
 determining $2t=T+\bar T$ as a function of $({\cal R}, b, z^i)$. If $V_c$ is a quadratic function of $t$, then the theory will be equivalent to an ${\cal R}^2$ theory coupled to matter. 

In supergravity $V_c$ is given in terms of the $Y$ function and the superpotential $W(\psi^I)$. In particular given the potential in the Einstein frame $V_E$, $V_c$ is given by
 $$
 V_E=e^K \left\{ \left(W_I+K_I W \right)  K^{I \bar J}\left(\bar W_{\bar J}+K_{\bar J} \bar W \right)-3\abs W\abs^2\right\}+ D{\rm - terms}
 $$
 \be
 V_c=Y^2\, V_E\, .
 \ee
Here $K$ is the K\"ahler potential (a real function of the scalars), which defines the metric $K_{I\bar J}$ on the scalar manifold via its holomorphic and anti-holomorphic derivatives:
  \be
  K_{I\bar J}={\partial \over \partial \psi^I} {\partial \over \partial \bar  \psi^{\bar I}}\, K(\psi^I, \bar \psi^{\bar J})\, ~~~{\rm with} ~~~ K=-3\log Y\, .
  \ee
  Hence, 
  \be
    K_{I\bar J}=-3{Y_{I\bar J}\over Y}+3{Y_{I}\,Y_{\bar J}\over Y^2}\, .
  \ee
 $K^{I\bar J}$ is the inverse of $K_{I\bar J}$.
 
The conformally equivalent Einstein frame action is given by
 $$
 S_E=\int d^4x\sqrt{\abs g \abs} ~\left[ {1\over 2} {\cal R}-K_{I\bar J}\,\partial_{\mu}\psi^I \partial^{\mu}{\bar \psi}^{\bar J}- V_E\right] ,
 $$
 \be
S_E =\int d^4x\sqrt{\abs g \abs} ~\left[ {1\over 2} {\cal R}-{3\over 4}{D_{\mu}D^{\mu}\over Y^2 }-{3\over 4}{J_{\mu}J^{\mu } \over Y^2}-3\left({-Y_{I\bar J}\over Y }\right)\,\partial_{\mu}\psi^I \partial^{\mu}{\bar \psi}^{\bar J}- V_E \right]\, ,
\ee
$~$\\
where the term quadratic in the current $J_{\mu}$ is induced by the auxiliary vector field $A_{\mu}$ of the supergravity multiplet, 
while the $D_{\mu}$ term is induced by the rescaling of the metric going from the Jordan to the Einstein frame:
\be
J_{\mu}=i(Y_{\bar I}\partial_{\mu} { \bar \psi}^{\bar I}-Y_{ I } \partial_{\mu}\psi^{ I}), ~~~~~~
D_{\mu}=Y_{\bar I}\partial_{\mu} { \bar \psi}^{\bar I}+Y_{ I } \partial_{\mu}\psi^{ I} \equiv\partial_{\mu}Y \, .
\ee
Like in the $SO(1,1+n)$ case (see eq.(\ref{scalarphi})) we introduce the no scale field $\phi$ as
\be  \alpha \phi=  \log Y  ~~{\with}~~~ \alpha= \sqrt{2\over 3}\, .
\ee
The $D_{\mu}$ kinetic part of the action becomes the kinetic term of the field $\phi$, and the action takes a very suggestive form: 
\be
S_E =\int d^4x\sqrt{\abs g \abs} ~\left[ {1\over 2} {\cal R}-{1\over 2}\partial_{\mu} \phi \partial^{\mu} \phi -{3\over 4}e^{-2\alpha\phi} J_{\mu}J^{\mu } - 3\, e^{-\alpha\phi} \partial_{\mu}z^{\i} \partial^{\mu}{\bar z}^{\bar \i}- V_E \right].
\ee

Some comments are in order:
\begin{itemize}
\item The kinetic part of the above supersymmetric extension defines a scalar manifold which is isomorphic to the $SU(1,1+n)$ manifold of the no-scale supergravity model \cite{sun1noscale}:
\be
K=-3\log\left(T+\bar T -\abs z^{i} \abs^2 \right)\, \longrightarrow~~~ {\rm \cal M}(T,z^{i})={SU(1,1+n)\over U(1)\times SU(1+n)}\, .
\ee
\item The Ricci tensor ${\rm \cal R}_{I \bar J}$ is proportional to the metric $K_{I \bar J}$ rendering the scalar manifold a maximally symmetric Einstein space: 
\be
{\rm \cal R}_{I \bar J}=-{\partial \over \partial \psi_I}{\partial \over \partial \bar \psi_{\bar J}} \log {\rm det}(K_{I\bar J})=-{\partial \over \partial \psi_I}{\partial \over \partial \bar \psi_{\bar J}} \, \log Y^{-(2+n) }=-{2+n \over 3} \, K_{I \bar J} \, .
\ee
The exponent $-(2+n)$ of $Y$ can be also derived, inspecting the expression of the action $S_E$ in terms of $\phi$, $J_{\mu}$
and $z^{i}$. A factor of $-2$ comes from the term $J^2_{\mu}=(2\partial_{\mu}b+\dots)^2$, which contains the kinetic term of the axion field $2b=i(\bar T-T)$ modulo $z^{i}$ fibrations. The additional factor $-n$ comes from the kinetic part of the $n$ complex scalars $z^{i}$.
\end{itemize}

The kinetic part of the $SU(1,1+n)$ theory is manifestly scale invariant, in a very similar way with the $SO(1,1+n)$ non-supersymmetric model. In the Einstein frame the scale symmetry 
acts as follows:
\be
T\rightarrow e^{2\sigma} T, ~~~~~z^{i}\rightarrow e^{\sigma} z^{i} ~~~\longrightarrow ~~~Y\rightarrow e^{2\sigma} Y,~~~e^{\alpha\phi}\rightarrow e^{2\sigma}e^{\alpha\phi}, ~~~b\rightarrow e^{2\sigma} b.
\ee
 The scale symmetry is an invariance of the potential and also of the gauge and fermionic sectors, provided that the resulting supergravity ${\bf G}$ function 
  \be
 {\bf G}\equiv K+\log\abs W\abs^2\, , ~~~~~~~K=-3\log Y \, ,
  \ee
  and the gauge kinetic holomorphic function $f_{ab}$ are invariant under the scale transformations defined above.
  Under these scale transformations $K\rightarrow K- 6\sigma$, imposing that $W$ has scaling weight 3: $W\rightarrow e^{3\sigma}\,W$. This
  drastically reduces the choices for the holomorphic functions $W$ and $f_{ab}$. Assuming the absence of fields with zero scaling weight (or else) 
and focusing to positive powers of fields in $f_{ab}$ and $W$, the possible choices
  are strongly restricted. Namely $f_{ab}$ is a constant, proportional to the inverse of the gauge coupling constant square:
  \be
  f_{ab} ={\delta_{ab} \over  g^2_{\alpha}}.
  \ee
  For simplicity we choose $ f_{ab} $ to be block diagonal in the gauge sector. This choice is justified in the absence of non-singlet fields (under the gauge group) with zero scaling weight. 
The possibility of field dependent couplings appearing in $ f_{ab}$, as in the heterotic superstring case where the coupling depends on the dilaton field $S$, 
will be discussed in section \ref{string}.
 
 Assuming for the moment that the $T$ field and $W$ are singlet under the gauge group -- we will return to this important point latter on -- insures that the $D$-part of the potential is scale invariant as well:
  \be
  V_D=\,{g^2_{\alpha}\over 2}  \, \left(D^{a}\right)^2\,=\,{ g^2_{\alpha}\over 2}~\left( K_i \, (T^{\alpha} )_{ \bar \i j}  \, z^{j} \right)^2={g^2_{\alpha}\over 2}~\left({ 3\,{\bar z^{\bar \i} \, (T^{\alpha} )_{\bar \i j} 
  \, z^{j} } \over Y}\right)^2\, ={g^2_{\alpha}\over 2}~\left({ 3\,{\bar z^{\bar \i} \, (T^{\alpha} )_{\bar \i j} 
  \, z^{j} } }\right)^2\,e^{-2\alpha\phi}.
  \ee
  As we already stated before, the $F$-part of the potential will be scale invariant once $W$ has scaling weight 3. 
This requirement leads to three possibilities (or linear combinations of those):\\
  \vskip0.1cm
  
 {\bf (i) $W=c\, T^{3/ 2}$, $~~~$ (ii)  $W=c_{ijk}\, z^{i}z^{j}z^{k}$, $~~~$ (iii)  $W=c_{k}\,z^{k}\,T\,$}.
  
  \subsubsection{Case (i). Anti de Sitter realization of the ${\cal R}^2$ theory } 
 The first choice, $W=c\, T^{3/ 2}$, gives rise to a scale invariant model with {\it negative} cosmological term. The vacuum corresponds to an  anti de Sitter ($AdS$) space. 
The parameter $\mu^2$ multiplying the ${\cal R}^2$ term turns out to be negative.  
 The minimal model assumes the absence of conformally coupled scalars, which means that the theory is conformally equivalent to a pure ${\cal R}^2$ supersymmetric theory. 

Indeed, the potential turns out to be negative:
 $$
 V_E=e^K\left({\abs K_T W+W_T \abs^2 \over K_{T\bar T}}-3\abs W \abs^2 \right)~=~
 {3\abs c\abs^2 \over 8 }\left({ \abs T\abs \, b^2 \over t^3 }  -  { \abs T\abs \,( t^2+b^2) \over t^3 } \right)\, ,~~~~~~~
 $$ 
 \be
~~~~V_E = - {3\abs c\abs^2 \over 8 }\sqrt{ 1+\left(b/t \right)^2 } ~~~~~{\rm and}~~~~
m_{3\over 2}^2=e^K\abs W \abs^2= {\abs c\abs^2 \over 8 }
\left( \sqrt{1+\left(b/ t \right)^2 }~\right)^{3} .
 \ee
 The potential is invariant under the scale transformation $T=t+ib \rightarrow e^{2\sigma} T $ as expected.
 The stationary point occurs at $b=0$. At this point, the positive part of the potential proportional to the $F$-auxiliary field vanishes, $F_T=e^{K\over 2}( K_T W+W_T )=0$, implying the preservation of supersymmetry. The potential saturates the Freedman bound:
 \be
 V_E=-3m^2_{3\over 2} \big \abs_{b=0}\, .
 \ee
 The classical vacuum corresponds to an $AdS$ space with non-vanishing gravitino mass term $m^2_{3\over 2}= {\abs c\abs^2 \over 8 }$ and $V=- {3\abs c\abs^2 \over 8 }$. 
 The parameter $\mu^2$ of the conformally equivalent $R^2$ theory is given by
 \be
 \mu^2=V=-3m^2_{3\over 2}\, .
 \ee
 
 \subsubsection{Case (ii). Flat space realization and the connection to no-scale models}
 The second choice is a particular case of a class of no scale models \cite{sun1noscale,noscale,FKZ} where the $Y$ function is linear in $T+\bar T$, while 
 the superpotential is an arbitrary function of $z^{i}$ but {\it  independent of $T$}:
 \be 
 Y=T+\bar T-g(z^{i}, \bar z^{\bar \j})\, ,~~~~~~W=W(z^{i}) \, .
 \ee
Here $g(z^{i}, \bar z^{\bar \j})$ is an arbitrary real function. What is special for this class of models, valid for any $g(z^{i}, \bar z^{\bar \j})$ and $W(z^{i})$, is that the potential is semi-positive definite, $V\ge 0$, independently if supersymmetry is preserved or broken at the vacuum. At any minimum of the potential where $V=0$, the $T$ direction is flat, while the gravitino mass term is undetermined at the classical level. This degeneracy of the vacuum is lifted at the quantum level of the theory as was shown more than 30 years ago \cite{sun1noscale,noscale}. 
 
We stress here that in order to lift the vacuum degeneracy at the classical level and thus stabilize the no-scale modulus $T$, it is necessary to determine, at a fundamental level, the origin of $T$-dependent terms (either in the superpotential or in $D$-terms). This was a long standing and tedious problem in supergravity and especially in the string induced no-scale models. It is well known by now that random modifications destabilize drastically the vacuum giving rise most of the time to anti de Sitter vacua with unbroken supersymmetry.  

The other fundamental property, valid for general $g(z^{i}, \bar z^{\bar \j})$ and $W(z^{i})$, is that the potential depends only through the derivatives of $W$, 
as in the global supersymmetric cases \cite{sun1noscale}
\be
V={3\over Y^2}\,  W_{i}\,  g^{ i \bar \j} \, {\bar W}_{\bar \j} +V_D\, .
\ee
This special property of this class of models shows that adding a constant term  $w$ to the superpotential   
\be
W \longrightarrow W+w\,  ,
\ee
leaves the scalar sector of the theory unchanged.  The gravitino mass and the fermion Majorana masses are shifted by $w$. 
Suppose the theory based on $W$ is supersymmetric at $<W>=0$. Then the theory with a shifted superpotential $W+w$ breaks supersymmetry. The non-zero gravitino mass is 
$m^2_{3\over 2}=\abs w\abs ^2 / Y^3$.

The  $SU(1,1+n)$ scale invariant theory is based on the choice $g=\abs z^{i}\abs^2 $ which is quadratic, with a trilinear superpotential ($w=0$). 
The choice $W=c_{ijk}\, z^{i}z^{j}z^{k}+~w$ {\it gives rise to scale invariant scalar sector even with non vanishing $w$}. The classical potential is semi-positive definite and quartic in the fields $z^{i}$, modulo the dressing coming from $\phi$ ($1/Y^2=e^{-2\alpha \phi}$): 
 \be
 V_E=e^{-2\alpha \phi} \left( z^4~{\rm terms} \right). 
 \ee
 The classical vacua of the model are characterized by $W_{i}=0$ and $D^a=0$ ($F$-flatness and $D$-flatness) with $V_E\equiv 0$. The no scale modulus $\phi$ and the gravitino mass remain undetermined at the classical level. This shows that in this class of models the $F$-part of the potential can never generate a non zero cosmological term. The only remaining way is via a contribution from non trivial Fayet-Iliopoulos $D$-terms. In this case the superpotential transforms 
 non-trivially under a local $U(1)$ $R$-symmetry. This possibility will be investigated in the subsequent sections.

 \subsubsection{Case (iii) de Sitter realization of the ${\cal R}^2$ theory } 
 In the third case the superpotential of the $SU(1, 1+n)$ model has a specific $T$ dependence, preserving however the scale symmetry due to a linear dressing by $z^{i}$. As we will see below the contribution of the $F$ part of the potential to the cosmological term is non trivial. However the $z^i$ directions are unstable. Here also a non-trivial contribution coming from a $U(1)_R$ Fayet-Iliopoulos $D$-term is necessary in order to stabilized the $z^i$ directions. 
 
 In order to simplify the presentation, we keep only one linear combination of the fields $z^i$, $hz\equiv c_{k}\,z^{k} $, together with the no-scale modulus $T$. 
The field $z$ is charged under the $U(1)_R$ local gauge symmetry while $T$ is neutral. Our analysis is reduced to the study of the $SU(1,2)$ model with a K\"alher potential $K$ 
and a superpotential $W$ given by:
 \be
 K=-3\log Y\, , ~~~~Y=T+\bar T-z\bar z\, ,~~~~W=h\,z\,T\, .
 \ee
 The metric of the scalar manifold $K_{I\bar J} $ is given by
 $$
 K_{T\bar T}={3\over Y^2}, ~~~~~~~~~~K_{T\bar z}={3\over Y^2}Y_{\bar z } =-{3\over Y^2}{z}~\, ,~~~~~~~~~~~~~
 $$
 \be
 K_{z\bar T} ={3\over Y^2}Y_z =-{3\over Y^2}{\bar z}, ~~~ K_{z\bar z}={3\over Y^2}(Y_z\, Y_{\bar z } -YY_{z\bar z})
 ={3\over Y^2}(T+\bar T)\, ,
 \ee
 which simplifies to
  $$
 K_{T\bar T}={3\over Y^2}\, , ~~~~~~~~~~~~~K_{T\bar z} =-{3\over Y^2}{z}\, ,~~~~~~~
 $$
 \be
 K_{z\bar T} =-{3\over Y^2}{\bar z}\, , ~~~~~~~~~ K_{z\bar z}={3\over Y^2}(T+\bar T)\, .
 \ee
 The inverse of the metric $ K^{I\bar J}$ is given by
 $$ 
  K^{T\bar T}={Y\over 3}(T+\bar T)\, ,~~~~~~~~~~~K^{T\bar z}= {Y\over 3} {\bar z} \, ,
 $$
 \be
K^{z\bar T}={Y\over 3} { z} \, ,~~~~~~~~~~~~~~~~~~ K^{z\bar z}={Y\over 3}\, .
 \ee

The scalar potential has the standard generic form of $N=1$ supergravity: 
$$
V_F={1\over Y^3}\left\{ {\left[W_I+ WK_J \right] K^{I\bar J} \left[ \bar W_{\bar J}+ \bar  W K_{\bar J} \right] -3\abs W\abs^2 }\right\}\, ,
$$
\be
K_T =-{3\over Y}, ~~~K_z={3\over Y}{\bar z},  ~~~W_T=hz,~~~W_{z}=hT.
\ee 
With the above choice for $W$, the $F$ part of the potential becomes
\be
V_F={\abs h\abs^2 \over 3Y^2}\left\{ \abs T\abs^2 -(T+\bar T) \abs z\abs^2\right\}
={\abs h\abs^2 \over 3}\left\{ {b^2+t^2  -2t \abs z\abs^2 \over Y^2}\right\}\, .
 \ee
It appears to be unstable in the $z$ direction. To study further this instability, we utilize the variable $Y=e^{\alpha\phi}$ rather than $t$:
 \be
  V_F={\, \abs h\abs^2 \over 12}\left\{ 1+4{b^2\over Y^2} -2 {\abs z\abs^2 \over Y}-3{\abs z\abs^4\over Y^2}  \right \} \, .
\ee
Introducing the scale invariant fields 
\be
B={b\over Y}\, ,  ~~~~~~\Phi= {~z \over \sqrt{Y}}\, ,
\ee
the $F$ part of the potential simplifies to a manifestly scale invariant form:
\be
  V_F={\, \abs h\abs^2 \over 12}\left\{ 1+4{B^2} -2 {\abs \Phi\abs^2 }-3{\abs \Phi\abs^4}  \right \}.
\ee

At the extremum with $B=\Phi =0$ the potential becomes constant with a positive cosmological term. At this extremum the no-scale modulus $\phi$ is a flat direction since it is only derivatively coupled, and thus invariant under translations by a constant. The $B$ field direction is a minimum with a mass  proportional to the cosmological constant. The $\Phi$ direction, however, has a maximum giving rise to instabilities. Even worse the potential in this direction is unbounded from below. In order to bypass this pathological behavior some modifications were proposed in the literature. 
 In one the  K\"alher function gets modified by suitably changing the $Y$ function of the initial $SU(1,2)$ scale invariant model 
 with $g(z,\bar z)=\abs z\abs^2$.  The change amounts to adding higher order terms to 
 $\abs z\abs^2$ \cite{Cecotti,N=1susyStarobinsky,DtermsOnly,NewFerraraPorrati}:
 \be
 Y=(T+\bar T) -\abs z\abs^2 +\zeta \abs z\abs^4+\dots \, ,
 \ee
 where $\zeta(T, \bar T)$ can be in general $T$ dependent. This modification can stabilize $z$ 
 at $z=0$ at least locally. Another proposal is that of Ref. \cite{Antoniadis} where 
 the field $z$ is assumed to be a composite field, namely a fermion bilinear of the goldstinos, so that $z^2=0$. 
 We will not consider these cases since our motivation goes beyond finding a Starobinsky-like model in supergravity. In particular, we would like to find such a model 
in the context of a more fundamental theory, like for instance superstring theory. So, we have to maintain the symmetry structure of the theory (here the $SU(1,n+1)$ scale invariant structure).
 
 As we show in the next section, the classical vacuum of the theory is de Sitter space {\it  if
 and only if } there is a non-trivial contribution from a non trivial $U(1)_R$ symmetry $D$-term necessary for the stabilization of the $z^i$ fields. 
Under this particular $U(1)_R$, the superpotential transforms with a non-trivial phase giving rise to a non-zero Fayet-Iliopoulos term (FI-term).

%%%%%%%%%%%%%%%%%%%%%%%%%%%%%%%%%%%%
%%%%%%%%%%%%%%%%%%%%%%%%%%%%%%%%%%%%
\section{De Sitter realization with Fayet-Iliopoulos $D$-terms}
\label{anomaly}
 Focusing on the de Sitter cases, 
 we must bypass two possible obstructions: the first is associated with the instabilities appearing in the $F$ part of the potential due to a $T$-dependent superpotential like for instance the one of case $(iii)$.  The second obstruction is associated with the flatness of the 
$F$ part of the potential due to the ``no-scale'' structure of the theory, as described for instance in case 
$(ii)$.  We will show below that both obstructions can be resolved by the contribution of a FI $D$-term originating from a $U(1)_R$ symmetry.  
We strongly believe, even if we do not have a general proof, that the precise $D$-term resolution is generic for all the scale invariant effective supergravities induced by superstrings.
(See section \ref{string} for more details.)  
 
 \subsection{$W=h\,z\,T$ with a $U(1)$ $D$-term stabilization of de Sitter }
\label{linearstab}
 
As we showed in the previous subsection, the $F$ part of the potential is unstable. This is due to non trivial dependence on $T$ of the
superpotential. The solution we propose in order to stabilize the classical de Sitter background is via generating a non-trivial contribution to the potential 
from a local $U(1)$ gauge symmetry. The choice of the $U(1)$ gauging must be compatible 
with the isometries of the scalar manifold defined by the K\"ahler potential. Also in case 
the superpotential transforms non-trivially under this $U(1)$, the ${\bf G}$ function of $N=1$,
supergravity must remain invariant: 
\be
{\bf G}=K-\log \abs W\abs^2 \, ,~~~~~~~~~W=h\, zT,~~~~K=-3\log \{T+\bar T -\abs z\abs^2\}.
\ee
 There are two available choices:
 \be
( a)~U(1)_R\, : \, z\to e^{iw} z, ~~~~~~~(b) ~ O(1,1)_d :~   z \to e^{\sigma} z~~ {\rm and}~~T\to e^{2\sigma} T.
 \ee 
The third possible isometry of $K$ which shifts $T$  
$$
(c) ~R_t\,: ~T\to T+ is \, ,
$$ 
is incompatible with a linear in $T$ superpotential. 
In both permitted cases, $U(1)_R$ and $O(1,1)_d$, the superpotential transforms non trivially, 
\be
~U(1)_R\, : \, W \to e^{iw} W, ~~~~~O(1,1)_d\, :~   W \to e^{3\sigma} W~.
\ee
Therefore, in both cases there is a non-trivial contribution to the $D$-terms coming from the superpotential. This contribution is nothing else but the Fayet-Iliopoulos 
term realized in supergravity theories, in the old minimal formulation \cite{MinimalSUCRA}.
We will consider below both cases since both of them correct the pathological behavior of the $F$ part of the potential.

\subsubsection{$U(1)_R$ gauging}
We first consider the $U(1)_R$ gauging. The associated $D$-term becomes
\be
~U(1)_R\, \longrightarrow D_R={\bf G}_i (qz)^i=q\left(K_z z+{W_z z\over W} \right)= q\left({3\abs z\abs^2\over  Y} +1\right)\, .
\ee
In terms of the scale invariant field $\Phi=z/Y^{1/2}$, it simplifies to
\be
D_R=q\left(1+3\abs\Phi\abs^2\right) \, ,
\ee
where q stands for the charge unit of $U(1)_R$. The first term in the equation above is nothing but the Fayet-Iliopoullos term, 
originating from the non-trivial transformation of $W$. Combining the $F$ and the $D$ parts of the potential, we obtain 
  \be
  V={\abs h\abs^2 \over 12}\left\{1+4B^2-2\abs \Phi\abs^2-3\abs \Phi\abs^4\right\} +
  {g^2 q^2 \over 2}  \left( 1+3\abs \Phi\abs^2 \right)^2\, .
  \ee
For convenience we define the effective $F$ and $D$ couplings, $H^2$ and $G^2$, as:
   $$
   H^2={\abs h\abs^2 \over 12}, ~~~~~~~G^2=3\, {g^2 q^2 \over 2}\, ,
  $$
  \be
  V= \{ H^2+{1\over 3}G^2\} +4H^2\, B^2 + (G^2-H^2)\left\{2 \abs \Phi\abs^2+3\abs \Phi\abs^4\right\}. 
  \ee
The combined potential is suggestive. First, $V$ is manifestly scale invariant since it is expressed in terms of the scale invariant fields $B$ and $\Phi$. 
The field $\Phi$ is stabilized once 
  $G^2>H^2$. Both the $B$ and $\Phi$ mass terms are non tachyonic; so these fields are stabilized at $B=\Phi=0$. 
In this vacuum the $U(1)_R$ symmetry remains unbroken with vanishing gravitino mass term $m_{3/2}=0$. The field direction $Y=e^{\alpha \phi}$ remains flat.  
  The cosmological constant term,
  \be
  {\mu^2 }= H^2+{1\over 3}G^2 \, ,
  \ee 
 receives contributions from both the $F$ and $D$ terms of the potential. The contribution of the $D$ term is just that of the Fayet-Iliopoullos term.  

For $G^2<H^2$ the potential has pathological behavior as in the ungauged theory. The limiting case with
 \be
  G^2=H^2 \, , 
 \ee
 is also interesting. In that case $\Phi$ decouples from the potential and becomes a flat direction. The $R$-symmetry is spontaneously broken as soon as $\Phi \ne 0$. 
More importantly, the gravitino mass term is not vanishing as well: 
\be
m_{3/2}^2=e^{K} \abs W\abs^2\Big\abs_{B=0} =\abs h\abs^2 \, \abs \Phi\abs^2\, \left({2T_R \over 2Y}\right)^2= 3 H^2 \abs \Phi\abs^2(1+\abs \Phi\abs^2).
\ee
To our knowledge, this is the first realization of {\it ``no-scale models in a de Sitter"} background. 
 
 \subsubsection{$U(1)_d$ gauging}
In order to study the other possible gauging, it is convenient to perform some analytic field redefinitions of $T$ and $z$. 
We may transform back to the $T$ and $z$ frame when necessary. Firstly we rewrite ${\bf G}$ as follows,
 $$
 {\bf G}=-3\log \left (T+\bar T-\abs z\abs^2\right) +\log \abs h\abs^2  \left ( \abs z\abs^2 \abs T\abs^2 \right) 
 $$
 \be
 =-3\log \left\{ \left({T\over \bar T}\right)^{1\over 2}+ \left({\bar T\over T}\right)^{1\over 2}
 -{~\abs z \abs^2 \over  \abs T\abs}     \right\}+\log \abs h\abs^2  \left({~\abs z \abs^2\over \abs T\abs}\right),
 \ee
which exemplifies the scale invariance of ${\bf G}$. Secondly we perform the following analytic field redefinitions:
\be
T=e^{2i\psi}, ~~~~ \bar T=e^{-2i\bar \psi}, ~~~{\rm and}~~~\Phi_n ={z\, e^{-i\psi}} \, .
\ee
In terms of $\psi$ and $\Phi _n$, ${\bf G}$ becomes:
\be 
{\bf G}=-3\log \left\{ 2 \, {\rm cos}(\psi+\bar \psi) - \abs \Phi_n\abs^2\right\}+\log \abs h\,\Phi_n\abs^2\, .
\ee
The above expression manifests the symmetries of the theory.  One is the $U(1)_R$ symmetry acting on $\Phi_n$: $\Phi_n\rightarrow e^{iw}\Phi_n$. 
This is equivalent to the previous case. There are however two other inequivalent cases. The first consists of gauging the imaginary translations of $\psi$ without acting on $\Phi_n$.
We disregard this possibility since it cannot resolve the pathological behavior of the $F$ part of the potential. 
The remaining gauging involves both the imaginary translations on $\psi$ and the $U(1)_R$ transformation of $\Phi_n$. Notice that in this case, 
there is a contribution from the superpotential which generates a non trivial Fayet-Iliopoullos term. The associated $D$-term becomes
\be 
D_d= q\left(3\,  {2\,\xi\, {\rm sin}(\psi+\bar \psi)+\abs \Phi_n \abs^2  \over  2 \, {\rm cos} (\psi+\bar \psi) - \abs \Phi_n\abs^2  } +1\right) \, , 
\ee
where $\xi q$ is the relative charge associated to the $\psi$ imaginary translation.  In terms of the initial $T$, $z$ variables this becomes
\be 
D_d= q\left(6\,\xi B\, + 3\,\abs \Phi \abs^2  +1\right) \, .
\ee

The difference from the $U(1)_R$ gauging is the presence of $B$ field, which appears linearly in the $D$ term and whose coupling is given in terms the FI-parameter $\xi$.
This is a fundamental difference, since in the absence of the $F$ term contribution, the $D$ term can be zero when
$6\xi B=-(3\,\abs \Phi \abs^2 +1)$. 
The difference however has an even more conceptual significance, as it is related to the anomaly of the $U(1)_R$ gauge symmetry. 
In fact, the origin of a non-vanishing FI-parameter $\xi$ is due to quantum effects, necessary for the cancelation of  the anomaly of the $U(1)_R$ gauge symmetry, as we will discuss in the following.
Indeed, when $W$ transforms under a $U(1)_R$ , this symmetry is anomalous. As soon as this symmetry is not gauged there is no obstruction. However, when this symmetry is gauged then {\it the theory is inconsistent } unless a mechanism, which resolves this anomaly exists. The most popular solution is via an axion field or via the Green-Schwarz mechanism. In both cases the anomaly is canceled by gauging the translation of an axion-like field, simultaneously with the initial anomalous $U(1)_R$. This is the difference between the ``anomalous''  classical gauging of $U(1)_R$ and 
the ``non-anomalous'' gauging based on $U(1)_d$. In the first case $D_R$ never vanishes, while in the second case $D_d$ can vanish. The axion field in the second case is nothing but Im$\, \psi$. Our claim is that at the classical level both the $U(1)_R$ and $U(1)_d$ gaugings are acceptable. However, at the quantum level only the $U(1)_d$ gauging provides a consistent anomaly free theory.  
Equivalently, at the quantum level the $U(1)_R$ gauging together with the ``axion-mechanism'' becomes identical to the $U(1)_d$ gauging. 

It is therefore necessary to examine in more detail the non anomalous $U(1)_d$ gauging. The potential in that case becomes
\be
  V=H^2\left\{1+4B^2-2\abs \Phi\abs^2-3\abs \Phi\abs^4\right\} +
  {G^2\over 3}  \left( 1+3\abs \Phi\abs^2+6\xi B \right)^2\, ,
  \ee
and so when $\xi\ne 0$ the trivial minimum $B=0$ is not the correct one. Minimizing with 
 respect to $B$, keeping $\Phi$ arbitrary,
\be
V,_{B}=8H^2\, B+4\xi G^2 (1+3\abs \Phi\abs^2+6\xi B)=0, ~\longrightarrow~ <B>=-{\xi G^2\over 2H^2 +6\xi^2 G^2}(1+3\abs \Phi\abs^2)\, ,
\ee
the effective potential for $\Phi$ becomes
\be
V=H^2\left\{1-2\abs \Phi\abs^2-3\abs \Phi\abs^4\right\} +{{\hat G}^2\over 3} (1+3\abs \Phi\abs^2)^2\, .
\ee
The effective coupling ${\hat G}^2$ and $<B>$ are given in terms of $H^2$ and $G^2$ by 
  $$
{\hat G}^2\,= \left({ H^2G^2 \over H^2 +3\xi^2 G^2}\right),~~~~~{\rm and}~~~~ 2H^2<B>=- {\hat G}^2\, (1+3\abs \Phi\abs^2)\, .
  $$
 In terms of ${\hat G}^2$ , the effective potential has a similar form as in the $U(1)_R$ case: 
  \be
  V= \left\{ H^2+{1\over 3}{\hat G}^2\right\} + ({\hat G}^2-H^2)\Big\{2 \abs \Phi\abs^2+3\abs \Phi\abs^4\Big\}. 
  \ee
When $({\hat G}^2-H^2)>0$ or  $G^2(1-3\xi^2)>H^2$, $\Phi$ is stabilized at $\Phi=0$. 
 If $({\hat G}^2-H^2)=0$, $\Phi$ becomes a flat direction with non-zero gravitino mass term (de Sitter no-scale model). 
The case $\xi=0$ reproduces a pure $U(1)_R$ gauging. The charge $\xi$ is restricted to be $\xi^2 <1/3$, otherwise the potential is unstable in the $\Phi$ direction. The precise value of $\xi$ is expected to be fixed by the anomaly cancellation condition at the quantum level of the model.   

In both examples of gauging there is non trivial contribution to the cosmological constant from both the $F$ and $D$ parts of the potential.  The $F$ part alone gives rise to instabilities. This however is not generic and depends crucially on the choice of the superpotential, as we show in the following subsection.

 \subsection{$W=c_{ijk} \, z^iz^iz^k$ with a $U(1)$ $D$-term lifting to de Sitter }
 
In this case $W$ is homogeneous of degree 3 in the $z^i$'s. The scale symmetry of the theory is manifest. The $F$ part of the potential, without a non-trivial contribution from a Fayet-Iliopoulos 
$D$ term, is semi-positive definite and therefore all possible vacua have vanishing potential ($V_F=0$).
 There are however three possible $U(1)$ gaugings which can lift the vacuum to de Sitter space. 
In contrast to the previous case, here there is no $T$ dependence in the 
 superpotential. The three isometries of $K$ are 
  \be
( a)~U(1)_R\, : \, z\to e^{iw} z, ~~~~(b) ~ O(1,1)_d :~   z \to e^{\sigma} z~~ {\rm and}~~T\to e^{2\sigma} T ~~~~~(c) ~R_t\,: ~T\to T+ is \, .
\ee
We will examine all possible gaugings in what follows. 

\subsubsection{ $U(1)_R$ gauging} 

 Since $W$ transforms non trivially under $U(1)_R$, it generates a non-trivial Fayet-Iliopoulos term while the $F$ part of the model is quartic in the $z^i$'s:
\be
U(1)_R\, : ~~z^i\rightarrow e^{iw}z^i,~~~~{\rm and}~~W\rightarrow e^{3iw}\,W\, .
\ee 
Keeping only one combination of the fields for simplicity, 
\be
W=c_{ijk}z^i z^jz^k =h\, Z^3, ~~~~V_F={3\abs h\abs^2}{\abs Z\abs^4\over Y^2},~~{\rm and} ~~
V_D={ g^2\over 2} 9\, q^2 \left( {\abs Z\abs^2\over Y}+1\right)^2,
 \ee
 so that $V$ becomes (in terms of the scale invariant field $\Phi=Z/Y^{1\over 2}$), 
 \be
 V_R=H^2 \, \abs \Phi\abs^4 +G^2\left(1+ \abs \Phi\abs^2 \right)^2,~~~~~~~~~~\left(H^2=3\abs h\abs^2,~~G^2=9{ g^2 \, q^2\over 2}  \right)\, .
 \ee
 The theory and in particular the potential are manifestly scale invariant as expected. There are no instabilities in both the $F$ and $D$ parts of the potential. The stationary point $\Phi=0$ is always a minimum and the vacuum solution is de Sitter space. The $\phi$ field remains massless, and since $\Phi=0$ the gravitino mass term is zero with the $R$-symmetry unbroken. Contrary to the previous example with linear $T$-dependence in $W$ ( case (iii)), in this model there is no contribution to the breaking of supersymmetry from the $F$ term. The breaking is purely a $D$ breaking. 
 
There are other possible gaugings of this cubic model. Namely, the $U(1)_d$ gauging where we gauge the dilatation symmetry together with the $R$-symmetry, as we did previously in case (iii). There is an additional possibility which was not present previously, namely gauging the imaginary translation of $T$ together with $U(1)_R$. As we will see below the vacuum structure changes drastically.
 
\subsubsection{ $U(1)_d$ gauging}   

 As in case (iii), the study of this gauging is more convenient in the ($\psi\,, \Phi_n$) representation of the fields, where the ${\bf G}$ function takes the form:
\be
{\bf G}=-3\log \left\{ 2 \, {\rm cos}(\psi+\bar \psi) - \abs \Phi_n\abs^2\right\}+\log \abs h\abs^2 \abs\Phi_n\abs^6\, .
\ee
Under $U(1)_d$, $\psi$, $\Phi_n$ and the superpotential transform as follows 
\be
U(1)_d\, :~~~\psi\rightarrow  i\xi qw\, ,~~~~~\Phi_n\rightarrow e^{iqw}\Phi_n, ~~~~~~W\rightarrow e^{3iqw}W\, .
\ee
In this representation the dilatation is expressed in terms of imaginary translations in $\psi$. The difference from case (iii) is that in this case
 the superpotential is cubic in $\Phi_n$ and not linear. The associated $D_d$ term is:
 \be 
D_d= 3q\left(\,  {2\,\xi\, {\rm sin}(\psi+\bar \psi)+\abs \Phi_n \abs^2  \over  2 \, {\rm cos} (\psi+\bar \psi) - \abs \Phi_n\abs^2  } +1\right) \, ,
\ee
where $\xi q$ is the relative charge associated to the $\psi$ imaginary translation.  
In terms of the initial $T$, $Z$ variables this becomes:
\be 
D_d= 3q\left(2\,\xi B\, + \,\abs \Phi \abs^2  +1\right) \, , ~~~~~
\left(\Phi={Z~\over Y^{1\over 2}}\right)\, .
\ee
Here also the difference from the $U(1)_R$ gauging is the presence of linear dependence in the $B$ field, proportional to the FI-parameter $\xi$. In this cubic case however
the $F$ part of the potential is independent of $B$. This is a fundamental difference since the $D$-term can be zero when
$2\xi B=-(\,\abs \Phi \abs^2 +1)$, which for $\xi\ne0$ screens the Fayet-Iliopoulos term. The total potential is semi-positive definite and becomes zero at the global minimum,
\be
 V_d=H^2 \, \abs \Phi\abs^4 +G^2\left(1+ \abs \Phi\abs^2 +2\xi B \right)^2\, \rightarrow\, <V>=0\, ~{\rm when} ~
 <\abs \Phi\abs^2>=0 \, \, {\rm and}\, < 2\xi B>=-1.
  \ee
The vacuum is flat space with $<\Phi>$=0. The supersymmetry, the $R$-symmetry and the scale symmetry are unbroken.  
The $B$ modulus is fixed while the $\phi$ field remains a flat direction with zero gravitino mass. 

The off-shell structure with $D$ flatness gives rise to a quadratic potential: 
$$
V=H^2\,  (2\xi B+1)^2, ~~~~~~{\rm with}~~~~\abs \Phi\abs^2=-(2\xi B+1)
$$
which can give rise to  chaotic inflation \cite{chaoticInflation} induced by the imaginary component of $T$,
very similar to that of Ref. \cite{ImaginarySUGRA}. Generically, the off-shell initial conditions for both $\Phi$ and $B$ give rise to two field ``chaotic inflation'',  
quartic in $\Phi$ and quadratic in $B$. 
In case the preliminary results of the BICEP2 collaboration are confirmed with $r\sim0.2$, 
then this class of supergravity realizations of two field chaotic inflation will become more attractive.

\subsection{$N=1$ supergravity with  scale violating terms and \\the emergence of inflationary potential: $U(1)_t$ gauging} 

The final allowable gauging involves imaginary translations in $T$.
The  $U(1)_t$ transformations are defined by their action on $T$ and $Z$ as follows
\be
U(1)_t\, : ~~~ T\rightarrow T+i\xi qw\, , ~~~~~~~Z\rightarrow e^{iqw}Z\,.
\ee
The fact that $W$ transforms non-trivially generates a Fayet-Iliopoulos term with corresponding FI-parameter $\xi$: 
\be
D_t=3q\left( {\abs Z\abs^2\over Y} -{\xi\over Y} +1  \right) =3q\left( {\abs \Phi\abs^2} -{\xi\over Y} +1   \right)\, .
\ee
So the potential becomes
\be
\label{Vt}
V_t=H^2 \, \abs \Phi\abs^4 +G^2\left( \abs \Phi\abs^2+1-\xi e^{-\alpha\phi}\, \right)^2 \, .
\ee
The analysis of the vacuum structure is identical to the $SO(1,1+n)$ model, in the non-degenerate case, where the scale breaking term $\xi/Y$=$\xi e^{-\alpha\phi}$ was introduced by hand.
In the supersymmetric case however it is induced after gauging of the non-anomalous $U(1)_t$, which involves gauging the axion shift symmetry corresponding to the imaginary translation of $T$,  $~2b\rightarrow 2b+\xi qw$. The extension of the gauging, 
$U(1)_R \to U(1)_t $ (or $U(1)_R \to U(1)_d $), is not an arbitrary choice, but {\it a necessary anomaly free consistency condition}, which must be valid at the quantum level of the theory.  

As was the case of the $U(1)_d$ gauging, the vacuum structure of the potential changes drastically as compared to the $U(1)_R$ anomalous gauging. In both 
$U(1)_d$ and $U(1)_t$ gaugings, the potential is semi-positive definite defining a global minimum with $V_{d,t}=0$. There is however a fundamental difference: in the $U(1)_d$ case the screening of the cosmological term is achieved via a scale invariant manner, involving the $B$-field, while in the $U(1)_t$ case the screening is due to the scale violating term,  $\xi e^{-\alpha \phi}$  $(\xi>0)$, involving the field $\phi$.  

$V_t$ has a structure similar to the Starobinsky potential. The scale violating term induced by the gauging, $ \xi e^{-\alpha \phi}$, is nothing but the the Einstein term $\xi {\cal R}$ of the Jordan frame. Therefore, gauging the non anomalous $U(1)_t$ with FI-term $\xi$ breaks the scale symmetry, by introducing a non-trivial Einstein term in the initially pure ${\cal R}^2$ theory. Equivalently, the initial $U(1)_R$ that creates a non-trivial cosmological term is anomalous. Including the axion shift in the local transformation, thus realizing the $U(1)_t$ gauging, resolves the anomaly, inducing a non scale invariant contribution. This contribution amounts to adding in the Jordan frame a non trivial Einstein term multiplied by the anomaly coefficient $\xi$: 
  \be
 {\cal R}^2\longrightarrow {\cal R}^2+\xi {\cal R},~~~~~~ \xi={\rm Tr }\,Q_R
 \ee
where $Q_R$ is the charge operator associated with the anomalous $U(1)_R$. Including the contribution of the charge of the axion shift, the anomalous $U(1)_R$ is extended to the non-anomalous $U(1)_t$. The global vacuum of the theory is now flat space due to the screening of the cosmological term by $\xi e^{-\alpha\phi}$. 

When several fields $\Phi_i$ are involved, they create a non-trivial scale invariant contribution to $V_F$, which plays the role of $V_c$ 
in the 
$SO(1,1+n)$ non-supersymmetric model we analyzed in detail in section \ref{soR}. Here the radial variable is $ \rho^2=\abs\Phi_i \abs^2$.   The non-degenerate case (which is similar in structure with the one field case in Eq.(\ref{Vt})) gives rise to the Starobinsky potential, once we integrate
over the radial field $\rho$ in the scale breaking era with $(\xi e^{-\alpha \phi} -1)>0$. In this era 
$\rho^2=(\xi e^{-\alpha\phi} -1)$ acquires a non trivial expectation value. In the inflationary era,
where $(1- \xi e^{-\alpha\phi}) >0$, the radial field is stabilized at $\rho=0$.  

When there are flat directions, $V_F$ can be set to zero along these directions and the total off-shell potential is fully controlled by $V_{D}$ :
\be
V=G^2 \left\{\rho^2 -(e^{-\alpha \phi}-1)\right\}^2\, .
\ee
As we already observed in the case of the $SO(1,1+n)$ model, the vacuum structure differs drastically from that of the Starobinsky model. The explicit presence of $\rho$, which is not stabilized by $V_F$, screens the exponential behavior of $\phi$ along the flat direction $\rho^2 =(e^{-\alpha \phi}-1)$ with vanishing total potential.  The effects of additional scale breaking terms, like for instance Coleman-Weinberg mass terms, will play an important role, since in the effective quantum potential of the theory both $\rho$ and $\phi$  will be fixed. The difference from the non-supersymmetric model is that the corrections are pretty much under control thanks to supersymmetry.

\section{String induced ${\cal R}^2$ models}
\label{string}

The scope of this section is to establish the appearance of the scale symmetry in superstring effective supergravity theories, which are treated at the semiclassical level ($\alpha'$-expansion). The origin of the symmetry is purely geometrical, and follows from the initial $10$-dimensional structure of the theory. The dilaton field  $\phi_{10}$ and the antisymmetric tensor field $B_{\mu\nu}$ will play a role.
 %\cite{StringDualities}. 
 To simplify the presentation we consider the cases where the $6\times6$ internal metric $g_{IJ}$ is block diagonal, given in terms of three  $2\times2$ blocks $g^{A}_{ij} , \, i,j=1,2, ~A=1,2,3$. As usual, we introduce the complex moduli fields $T^A$ and $U^A$
\be
T^A=\sqrt{ \det{g^A_{ij} }}+iB^A_{12}, ~~~~~~~~ U^A={  g^A_{11} +i g^A_{12} \over \sqrt{ \det{g^A_{ij}}}}\, ,
\ee 
as well as the complex four dimensional dilaton field $S$
\be
S=e^{-2\phi_{4}} +ia \, ,~~~~~~~e^{-2\phi_{4}} =e^{-2\phi_{10}}  \sqrt{ \det{g_{IJ} }}\,, \,~~~~  \partial_{\mu}a=
\epsilon_{\mu}^{\nu \rho \sigma}\, \partial_{\nu}B_{\rho \sigma}.
\ee 
 In terms of the volume moduli $T^A$, the complex structure moduli $U^A$ and the dilaton field $S$, the four dimensional effective action takes the following universal form:
\be
S=\int d^4x\sqrt{\abs g \abs} ~\left[ {1\over 2}{\cal R} -{\partial_{\mu} S \, \partial^{\mu}\bar S \over (S+\bar S)^2 }- {\partial_{\mu} T^A \, \partial^{\mu}\bar T^A \over (T^A+\bar T^A)^2}- 
 {\partial_{\mu} U^A \,  \partial^{\mu}\bar U^A \over (U^A+\bar U^A)^2 }  +\dots
\, -V_F-V_D \right]\, ,
\ee 
where the ellipses stand for the kinetic terms of all other excitations, which are under control in orbifold or Calabi-Yau compactifications from $10$ to $4$ dimensions. The above structure arises in all string compactifications
%\cite{StringDualities}\cite{StringStringDualities}\cite{FKZ}. 
utilizing some perturbative and non-perturbative equivalences between the known string theories, which exchange the role of the individual moduli. Namely: \\
$~$\\
(a) Mirror symmetry $T^A\leftrightarrow U^A$\, ,\\
(b) Heterotic-type IIA non-perturbative string duality: $ T^1\leftrightarrow S$\, ,\\
(c) Heterotic-type IIB non-perturbative string duality: $\, U^1\leftrightarrow S$\, ,\\
(d) Type IIA-type IIB orientifolds with D-branes and fluxes: $T^A \leftrightarrow U^A$, $\dots$\, .\\

%Although the above duality transformations seem to act trivially in the $S, \, T^A, U^A$ sector, they act non trivially in the other sectors of the theory.  

Many of these moduli fields can be frozen by turing on various fluxes within the different theories (see e.g. \cite{Fluxes} and \cite{Grana:2005jc} for a review on flux compactifications).
In the most symmetric cases, like for instance the $\Z_2\times \Z_2$ orbifold 
compactifications of the heterotic string or the $D_5$/$F_5$ orientifold constructions of the type IIB theory, none of the geometrical moduli $T^A$, $U^A$ is  
frozen.  However, in $\Z_3$ orbifold compactifications of the heterotic string, as well as in type IIB orientifold constructions with $D_3$ and $D_7$
branes, the $U^A$ moduli are frozen. Moreover, in type IIB compactifications the $S$ modulus can be frozen together with the complex structure moduli $U^A$. In the type IIA theory the situation is similar to the type IIB cases, as follows via the interchange of $U^A\leftrightarrow T^A$, equivalently via string-string dualities.
%\cite{StringDualities}\cite{StringStringDualities}
%(\cite{Fluxes}. 
An important result from the study of the type IIB cases is that all complex structure fields $U^A$, as well as the dilaton field $S$, are stabilized in generic orientifold compactifications with fluxes. In the string effective supergravities the fluxes are in one to one correspondence with the gaugings of the graviphotons, appearing in sub-sectors of the $N=1$ (or even $N=0$) theory with non-aligned extended supersymmetries $N>1$.

 In the heterotic cases the mechanism that stabilizes the $S$ modulus is more involved, since it incorporates non-perturbative effects, like for instance gaugino condensation \cite{gaugino} and fluxes\cite{Fluxes}.  At the level of the $N=1$ effective supergravity theory, these effects can be described in terms of suitable modifications of the superpotential, consistent with the string-string dualities,
%  \cite{StringDualities} \cite{StringStringDualities}-- 
  especially with the heterotic-type IIB orbifold/orientifold constructions.  A full understanding of the stabilization mechanism of $S$ in the heterotic cases has not yet been achieved. However, assuming that the non-perturbative string-string dualities with the type II cases hold, we may argue that such a mechanism is possible. 

Based on the above, we may assume that in both the type IIB and the heterotic effective supergravities, the complex structure moduli $U^A$ and the dilaton field $S$ are generically stabilized. The remaining moduli are the geometrical volume moduli $T^A$. Their kinetic terms are invariant under the scalings
\be
T^A \longrightarrow  e^{2\sigma_A}  \, T^A
\ee
and in particular, under the  diagonal scaling where $\sigma_A=\sigma$. This implies that the kinetic term of the diagonal direction $T=T^A$ appears always with a universal normalization coefficient: 
\be
- {\partial_{\mu} T^A \, \partial^{\mu}\bar T^A \over (T^A+\bar T^A)^2} \longrightarrow  -3\, {\partial_{\mu} T \, \partial^{\mu}\bar T \over (T+\bar T)^2}\, ,~~~{\rm with}~~~~K=-3\log (T+\bar T)\, ,
\ee
showing that $T$ is a member of an $SU(1,1)/U(1)\times U(1)$ manifold with a specific curvature\cite{sun1noscale,noscale,FKZ}. 
Performing a conformal transformation of the metric in order to pass to the Jordan frame, the field $T$ appears multiplying the Einstein term ${\cal R}$ without any kinetic term for $T+\bar T$: 
\be
\left[ {1\over 2} (T+\bar T)  \left({\cal R}+{2\over 3}A_{\mu}A^{\mu}\right) -J_{\mu}A^{\mu}\right] + \dots -(T+\bar T)^2 (V_F+ V_D)_E\, .
\ee  
This shows explicitly that once the potential in the Einstein frame is $T+\bar T$ independent, then the algebraic equation of the field $T+\bar T$ casts the theory to be conformally equivalent to an ${\cal R}^2$ theory.  

Switching on the other degrees of freedom associated with the off diagonal metric components, does not change this result. The diagonal $T+\bar T$ mode is related to the volume of the six dimensional compact manifold:
\be
\label{volume}
(T+\bar T)~\longrightarrow~ Y=e^{- {1\over 3} K(T^a,\, \bar T^a; \, z^I,\, \bar z^{\bar I})}\, .
\ee
$T^a$ stand for all components of the metric (the $h_{1,1}$ moduli fields) and the $z^I$ denote the Wilson line moduli, as well as the chiral matter fields arising from the orbifold twisted states in the heterotic string, and states localized on $D_3$ and $D_7$ branes in the Type II orientifold constructions.  
It is remarkable that $Y$ transforms homogeneously under the following scale transformations:
\be
T^a\rightarrow e^{2\sigma} \, T^a, ~~~ z^I \rightarrow e^{\sigma} \,  z^I, ~~~~ Y \rightarrow
e^{2\sigma}\,Y\, ~~ {\rm and }~~K \rightarrow K- 6\, \sigma \, .
\ee
States with scaling weight $ w={\bf 3}$, can also appear, like for instance the twisted singlet fields in the heterotic $Z_3$ orbifold compactification. Their presence however is compatible with the above transformation properties 
of $Y$ and $K$. The profound reason of this scale symmetry has its origin in the modular invariance of string theory\cite{FKZ}.  Remarkably, the above property of the geometrical $T$-sector extends to the $U$-sector, and to several $STU$-sectors thanks to the sting-string dualities.
It can be extended as well in $M$-theory and $F$-theory non-perturbative constructions. 
The other important property is that the classical scale symmetry extends to all sectors of the theory, and in particular to the $F$ and
$D$ parts of the potential. Equally significant for us is the presence of several $R$-symmetries 
which are associated to anomalous $U(1)$'s. These anomalous symmetries are always corrected at the string level via the axion gauging, as explained in the previous section in the framework of the $SU(1,1+n)$ effective supergravity.

In what follows we will present some stringy examples, generalizing  the
$SU(1,1+n)$ model to cases with more than one $T^A$ modulus, as appearing in $N=1$ heterotic orbifold compactifications and in type IIB orientifold constructions.   

\subsection{$(T_1, T_2,T_3)$ moduli with trilinear in $z$ superpotentials}

In typical situations the K\"ahler potential depends on the three $T_A$  main moduli as follows:
\be 
K=-\log Y_1-\log Y_2-\log Y_3, ~~~~~~Y_A=(T_A+\bar T_A) - \abs {z^{i }_A}\abs^2\, .
\ee
This structure emerges in both the heterotic and the type IIB cases after stabilizing the $U_A$ moduli as well as implementing 
the non-perturbative stabilization of the $S$ field. 
What is also typical is the dependence of the superpotential on chiral fields of scaling weight 1, via the following trilinear terms:
\be
W=d_{ijk}^{ABC}\, z^i_A\, z^j_B\, z^k_C\, .
\ee
To simplify the presentation we turn on only one such matter field per complex plane:
$$
W=h \, z_1\, z_2\, z_3\, .
$$
There are three distinct $U(1)^A_R$ $~R$-symmetries, one for each plane. Gauging  either
all of them, or one or two of them, or even the diagonal combination, induces Fayet-Iliopoulos terms. 
As we explained previously, these $U(1)^A_R$ symmetries are anomalous, and must be extended to  
either the $U(1)^A_d$ or the $U(1)^A_t$ gaugings, so that the resulting string effective theory
be anomaly free. The $U(1)^A_t$ gaugings introduce scale violating terms, which are responsible for an inflationary slow roll transition 
from a de Sitter to a flat Minkowski phase.

For the particular trilinear superpotential, the diagonal gauging yields the following {\it 3-field inflationary potential}: 
$$
V=\abs h\abs ^2   \Big(  \abs \Phi_2\,\Phi_3 \abs^2 + \abs \Phi_3\,\Phi_1 \abs^2 + \abs \Phi_1\,\Phi_3 \abs ^2  \Big) +
G^2 \Big( \abs\Phi_1\abs^2+ \abs\Phi_2\abs^2+  \abs\Phi_3\abs^2  +3-{\xi_1\over Y_1} -{\xi_2\over Y_2} -{\xi_3\over Y_3}\Big)^2\, .
$$
The slow roll inflationary transition becomes transparent once we express the $Y_A$'s in terms of the canonically normalized fields $\phi_A$:
\be
Y_A=e^{\gamma_A \phi_A}, ~~~{\gamma_A=\sqrt{2}}\, .
\ee
The over all no-scale modulus $\phi$ is nothing but the diagonal modulus (see Eq.(\ref{volume}))
\be
Y=\left({Y_1Y_2Y_3}\right)^{1\over 3}=e^{{\gamma \over 3}(\phi_1+\phi_2+\phi_3)}\, =\, e^{\alpha \phi}\, ,~~~~~~~\alpha=\sqrt{2\over 3}\, ,
\ee
as expected by the conformal equivalence with the pure ${\cal R}^2$ theory in the Jordan frame.

Several models can be constructed according to the anomalous $R$ gauging. In the framework of the effective supergravity this is just a choice. 
In string theory, the apparently anomalous $U(1)_R$ can be identified in 
any individual four dimensional model without any ambiguity.  From our experience with the 
$\Z_2\times \Z_2$ (asymmetric) orbifolds via the fermionic construction\cite{ChiralAndU(1)R}, at any time there is a net chirality emerging in a sector associated with one of the three $N=2$ complex planes, 
there is an anomalous $U(1)^A_R$. Therefore, {\it in any realistic string effective supergravity, the $U(1)^A_t$ non-anomalous gauging is not just a choice but a necessity}. 
Although the validity of this statement is automatic in the fermionic formulation (by construction), we believe that it is also valid in all consistent string compactifications.  
The above generic statement implies that at least a $U(1)^ A_t$ gauging is necessary for any sector which provides net chirality. If there is a symmetry between the three sectors, then 
the ``diagonal gauging'' described above suffices for the cancellation of the would be anomalous $U(1)^A_R$. In more general asymmetric constructions, it is necessary to consider individual non-anomalous $U(1)_t^A$ gaugings, for each sectors where chiral matter appears. The resulting potential of an anomaly free string effective supergravity theory becomes: 
\be
V=V_F+\sum_{A~ {\rm chiral}} G^2_A\Big( ~\abs \Phi_A^I \abs^2 +1-{\xi_A\,  e^{-\gamma \phi_A}} \Big)^2\, ,
\ee
where $V_F$ is the universal $F$ term contribution to the potential (which is independent of the 
$D$-term induced by the gauging). For the trilinear case this part is always scale invariant. 

The general analysis of section \ref{soR} can applied here. In the de Sitter and inflationary eras, where the $\xi_A$ scale violating effects are exponentially suppressed, 
the vacuum corresponds to an approximate de Sitter space. The $\Phi_A^I$ fluctuations are massive with their masses  protected by the induced de Sitter contribution even at the quantum level.  
If $V_F$ has no flat directions, there is a unique global supersymmetric flat vacuum with $\Phi_A^I=0$ and with all the $\phi_A$ moduli stabilized. In this case the structure of the of shell potential is similar to that of the simple Starobinsky model generalized for two and three inflaton fields. However, if  $V_F$ is flat in the $\Phi$ directions, 
then the flat vacuum turns out to be degenerate, with $D^A_t=0$, and with $\phi_A$ and $\Phi_A^I$ acquiring non trivial vacuum expectation values. At the quantum level the vacuum degeneracy will be lifted and both the $\phi_A$ and $\Phi_A^I$ vacuum expectation values will be determined, especially if supersymmetry is spontaneously broken.

The other non anomalous gaugings, $U(1)^A_d$, will not generate a slow roll inflationary era of the Starobinsky type. Although their relevance for inflationary cosmology may prove to be
important, in the context of the  ``new inflationary scenario'' or  ``chaotic inflation''\cite{chaoticInflation}, we do not analyze them 
further in this work.
 
\subsection{$(T_1, T_2,T_3)$ moduli with linear in $z$ superpotentials}

In this section we illustrate some stringy examples where fields with scaling weight ${\bf 1}$ appear linearly in the superpotential, generalizing similar models introduced by 
Cecotti \cite{Cecotti}  almost 20 years ago. 
In the type IIB theory this kind of superpotentials cannot arise in $(2,2)$ -superconformal compactifications (like Calabi-Yau compactifications). In the orbifold/orientifold constructions however, the obstructions can be easily  
bypassed if we introduce non trivial torsion. The simplest example involves Scherk-Schwarz  geometrical fluxes\cite{SS,SSstring}, which introduce  
the desired CY-torsions. These may break the supersymmetry.  In the $N=1$ supersymmetric heterotic cases, topological mass terms can arise in some of the $N=2$ sub-sectors of the theory\cite{FKZ}.  
In summary superstring constructions with geometrical fluxes permit linear superpotentials in both the heterotic and type IIB cases, even though the constructions are non trivial. 
Even at the classical supergravity level the generic $F$ part of the potential has pathological behavior. This pathology is due to the fact that the appearance of the $T$ moduli in the 
superpotential destroy in general its positivity properties due to the underline no scale structure. We showed explicitly how this pathological behavior is cured by suitable $D$-terms from gauging, lifting the tachyonic directions of the models -- see section \ref{linearstab}.   

We present three typical string inspired models based on linear superpotentials. 
The  K\"ahler potential is the same as in the case of the trilinear models described in the previous section:
\be 
K=-\log Y_1-\log Y_2-\log Y_3, ~~~~~~Y_A=(T_A+\bar T_A) - \abs {z^{i }_A}\abs^2\, .
\ee
The simplest case is when one such field, say  $z$=$z_1$, is turned on associated with the first complex plane. 
Then, there are two choices producing a scale invariant superpotential:
\be
W=h\,z\, (T_2-\epsilon T_3)\, , ~~~~{\rm with}~~~\epsilon=\pm1\, .
\ee
There are other interesting possibilities where the scale invariance is broken spontaneously, 
by the freezing mechanism which fixes $U^A_I\sim \xi \ne 0$ in the presence of non-trivial fluxes, both in type IIB orientifolds and also in the heterotic cases:
\be
W=h\,z (2T-\xi),~~{\rm for~instance~when} ~~T_2=T_3=T\, .
 \ee
The above model looks like the Cecotti model \cite{Cecotti};  however it differs from that as  it involves only two out of the three moduli.  
As we already emphasized before, the stringy origin of the above typical models is due to geometrical fluxes via a perturbative supersymmetric stringy Scherk-Schwarz ``mass generation'' mechanism. This must not be confused with the familiar Scherk-Schwarz mechanism that breaks supersymmetry.
 
\subsubsection{The linear $z_1$ model with supersymmetric, scale invariant, topological mass terms, $\epsilon$=$1$}

In this case the superpotential is $W=h\,z_1\, (T_2-T_3)$. The relative minus sign between $T_2$ and $T_3$ is of main importance, since for $T_2 \sim T_3$ it generates supersymmetric 
mass terms for the $z_1$ and $ (T_2-T_3)$ fluctuations. Furthermore the superpotential has two zero modes: $z_1=0$ and $(T_2-T_3)=0$. This is an essential difference with the $\epsilon=-1$ model 
we consider later. The $F$ part of the potential is well behaving, without instabilities in the $z_1$ direction: 
 \be
 V=\abs h \abs^2\left( {\abs T_2-T_3 \abs^2 \over Y_2 Y_3} +2 \abs \Phi\abs^2\right)_F +G^2\left( \abs \Phi\abs^2 +1 -{\xi_1 \over Y_1} -{ \xi \over Y_2} -{\xi \over Y_3} \right)^2_D  \, \, ,  
 \ee
where $\Phi=z_1/(Y_1)^{1\over2}$ is the scale invariant normalized field associated with the first complex plane. 
The $U(1)_t$ gauging takes into account the fact that the imaginary translations of $T_2$ has to be the same with that of $T_3$. This is the reason that the
in the $D$ term  $\xi_2=\xi_3=\xi$. The potential is scale invariant modulo the $\xi$-anomaly terms. It is also semi-positive definite and has a global flat vacuum:
\be
\Phi=0, ~~~\xi_1\, e^{-\gamma \phi_1}+\xi \, e^{-\gamma \phi_2}+\xi\, e^{-\gamma \phi_3}=1 ~~~{ \rm and ~also } ~~\phi_2=\phi_3.
\ee
From a cosmological point of view, this model can be analyzed as a two component inflationary model, since the potential attracts $\phi_3\to \phi_2$ rapidly. Indeed, 
\be
{\abs T_2-T_3 \abs^2 \over Y_2 Y_3}=(b_2-b_3)^2 \, e^{-\gamma (\phi_2+\phi_3)}+
{\rm sh}^2{\gamma\over 2}(\phi_2-\phi_3) \, ,
\ee 
or in terms of the scale-invariant fields $B_2, B_3$, 
$$
{\abs T_2-T_3 \abs^2 \over Y_2 Y_3}=\left(B_2\,e^{{\gamma\over 2}(\phi_2-\phi_3)} -B_3\,{e^{-{\gamma\over 2}(\phi_2-\phi_3)}}\right)^2+{\rm sh}^2{{\gamma \over 2} }(\phi_2-\phi_3).
$$
Thus the potential attracts exponentially $\phi_3\to \phi_2$ and $B_3\to B_2$.

\subsubsection{The linear $z_1$ model with tachyonic, scale invariant, topological mass terms, $\epsilon$=$-1$}

For $\epsilon=-1$ the $F$ part of the potential is unstable. The main reason is that the superpotential term can vanish only for $z_1=0$, since  $(T_2+T_3)\ne 0$. 
Therefore, this model creates tadpoles and a $D$-term is necessary to resolve the tachyonic structure of the $F$ part. This pathology is similar to the 
$F$-inflationary model of Cecotti \cite{Cecotti} and others. Can this pathology be cured  with an additional contribution from a $D$-term? 
As we show this can achieved only by the anomalous $U(1)_R$ gauging,
without the $\xi$ scale breaking terms arising in the $U(1)_t$ non anomalous gauging. Naively, the $(D_R)^2$
term stabilized the de Sitter phase, without any inflationary transition to the flat phase. In the $U(1)_t$ non anomalous case the potential becomes 
\be
 V=\abs h \abs^2\left( {\abs T_2+T_3 \abs^2 \over Y_2 Y_3} -2 \abs \Phi\abs^2\right)_F +G^2\left( \abs \Phi\abs^2 +1 -{\xi_1 \over Y_1} -{ \xi \over Y_2} +{\xi \over Y_3} \right)^2_D  \, \, , 
 \ee
 or
 $$
V=\abs h \abs^2 \left(B_2\,e^{{\gamma\over 2}(\phi_2-\phi_3)} +B_3\,{e^{-{\gamma\over 2}(\phi_2-\phi_3)}}\right)^2+\abs h \abs^2\, {\rm ch}^2{\gamma\over2}({\phi_2-\phi_3})\, 
$$
\be
-2\abs h \abs^2 \abs \Phi\abs^2+G^2\left( \abs \Phi\abs^2 +1 -{\xi_1 \over Y_1} -{ \xi \over Y_2} +{\xi \over Y_3} \right)^2  \, \, .
\ee
Examining the off-shell structure, we first observe that the two first terms are positive definite with a minimal contribution when $\phi_2=\phi_3$ and $B_2=-B_3$ $(T_2= \bar T_3)$. 
Along this direction the potential simplifies and becomes
\be
V_{\rm effective}=\abs h \abs^2-2\abs h \abs^2 \abs \Phi\abs^2+G^2\left( \abs \Phi\abs^2 +1 -{\xi_1 \over Y_1}  \right)^2  \, \, .
\ee 
At the classical level, where the $\xi_1$ term is absent, the model is well defined as soon as $G^2-\abs h \abs^2>0$. As soon as this bound is satisfied the $\Phi$ direction is massive, forcing the expectation value to be frozen at zero: $\Phi=0$. The vacuum is a de Sitter space with a cosmological constant $\mu^2=G^2+\abs h \abs^2$. However, when the breaking terms are swhiched on $\xi\ne 0$, the de Sitter vacuum is destabilized. Contrary to the previous case with $\epsilon=1$, where the global vacuum was flat Minkowski space, here the global vacuum has runaway behavior because of the tachyonic term $-2\abs h \abs^2\abs \Phi\abs^2$ coming from the superpotential. Most models of this type have similar behavior and have to be disregarded unless a stringy mechanism exists which can cure this  behavior. 
    
\subsubsection{The linear $z_1$ model with supersymmetric scale non-invariant mass terms}
As we already mentioned before, the origin of this class of models is due to the stabilization of the $U^I_A$ complex structure fields and also due to geometrical fluxes, 
which may break the scale invariance softly, introducing ``scale non-invariant mass terms'' in the superpotential: 
\be
W=h\, z_1\,(2T-\xi), ~~~~{\rm with}~~~T=T_2=T_3.
\ee
The $F$ part of the potential is by itself pathological, as in the initial Cecotti model \cite{Cecotti}. It is possible
here to gauge the non anomalous $U(1)^A_t$. 
In the Cecotti model there are no non anomalous $U(1)_t $ or $U(1)_d$ gaugings since all of them are broken by the choice of the superpotential.  
Only when $\xi=0$ the $U(1)_d$ gauging is possible that we have already treated, with a resulting  potential having cosmological applications in the framework of chaotic inflation. 
In the model we are proposing the non-anomalous $U(1)_t$ is realized in the first complex plane. Considering this gauging the potential becomes  
\be
V=\abs h \abs^2\left\{4 B^2 +\left(1-{\xi \over Y_2}\right)^2+\left({4\xi \over Y_2}-2\right)
 \abs \Phi\abs^2 \right\}_F+G^2\left( \abs \Phi\abs^2+1-{\xi_1 \over Y_1}    \right)_D^2.
\ee
In the de Sitter era, with both $Y_1=e^{\gamma\phi}$ and $Y_2=e^{\phi_2}$ large, the expectation values of the fields $B$ and $\Phi$ vanish, $<B>$= $<\Phi>$=$0$, 
as soon as  $G^2>\abs h \abs^2 $. There is a stable 
minimum with vanishing potential:
\be
\big<1-{\xi \over Y_2}\big>=0, ~~~  \big<1-{\xi_1 \over Y_1}\big>=0, ~~~ \big<B\big>=0, ~~~ \big<\Phi\big>=0~~{\rm with}~~\big<V\big>=0.  
\ee
The (mass)$^2$ term of $\Phi$ in the Minkowski vacuum is positive receiving contributions from both parts of the potential, while $B$ receives a mass from the $F$ part only:
$$
m^2_{\Phi}=2\abs h \abs^2+2G^2\, ,~~~~m^2_{B}=4\abs h \abs^2\, .
$$
When the scale breaking terms for both $\phi_1$  and $\phi_2$ are dominant, $\Phi$ is attracted to zero as in the de Sitter era. There are however {\it unstable directions} in the region where the scale breaking term
$Y_1$ is dominant while $Y_2$ subdominant. In this direction the expectation value of $\Phi$ can make the $D$-contribution vanish. Then
\be
V=\abs h \abs^2 \left(1-{\xi \over Y_2}\right)^2+\abs h \abs^2\left({4\xi \over Y_2}-2\right)\left( {\xi_1 \over Y_1}-1    \right)^2\, ,~~ \abs \Phi\abs^2={\xi_1 \over Y_1} -1>0\, .
\ee
Therefore, for ${4\xi /Y_2}<2$ the potential is unbounded from below when 
${\xi_1/ Y_1} $ is large . This pathology is harmless initially in the de Sitter era when the $\xi/ Y_2$ breaking term grows more rapidly than $\xi/Y_1$. 
The other possibility which avoids the asymmetric destabilization is by enforcing the diagonal direction via suitable geometrical flux terms.

\section{Conclusions}
\label{conclu}

 The first part of our work is motivated by the  observation that  the pure ${\cal R}^2$ theory is the only scale invariant gravity theory 
without ghost. As a first step we show that this theory is conformally equivalent to a conventional Einstein  gravity with an extra 
scalar degree of freedom $\phi$ without potential thanks to the underlying  classical scale invariance.  At the classical level there are three and only three possible vacua according to the sign of the coupling $\mu^2$ of the  ${\cal R}^2$ theory. The de Sitter vacuum ($dS$) arises, if  $\mu^2$ is positive,  the anti de Sitter vacuum ($AdS$), if is $\mu^2$ is negative  and the flat Minkowski space (${\cal M}$), if $\mu^2$ is zero. Although this is an interesting  observation it is not useful  in the absence of at least the observed Standard Model matter with the gauge, Yukawa  and self scalar interactions. This forces us to generalize the ${\cal R}^2$ theory by coupling it to conformal matter including all possible interactions. The resulting classical theory has the desired scale symmetry as a remnant of the conformal invariance of the matter sector.  
%At this second step level, the Standard Model of matter and interactions is successfully  coupled. 
Any mass breaking terms in both the bosonic and fermionic sectors are forbidden classically. We know however how the quantum corrections associated to the matter interactions violate the scale invariance inducing non-trivial scales due to renormalization effects, namely, introducing Coleman-Weinberg mass terms, anomalous dimensions in the fields and couplings which break ``softly" the classical scale invariance. This kind of scale violation effects are under control via the renormalizability of the matter sector around a weakly curved gravitational background (in the approximation where the gravitational background is treated classically). 

Disregarding at this step the well known field theory mass-scale hierarchy problem (which will not be present in the supersymmetric extension of the theory), we present in section 2 the classical ${\cal R}^2$ theory coupled to matter.  We show that in the Einstein frame the conformally coupled scalars and the dual field $\phi$ (the no scale modulus), associated to the extra degrees of freedom of the ${\cal R}^2$ theory, form together a non-trivial scalar manifold in the conventional Einstein frame: ${\cal M}_0 =H^{n+1}\equiv SO(1,1+n)/SO(1+n)$. This is an important result, since any extension (supersymmetric or not) of the scale invariant ${\cal R}^2$ theory has to contain ${\cal M}_0 $ as a submanifold. This universality  property of  ${\cal M}_0 $ let us to examine in more detail the vacuum structure of the universal sector based on $SO(1,1+n)$ theory in the presence of scaling violating terms, like for instance the undressed Einstein term ${\cal R}$ in the Jordan frame.  We have shown explicitly that the ${\cal R}$ term is the origin of an inflationary slow roll transition from the de Sitter to the flat space, in a very similar way as in the minimal Starobinsky model.  More importantly, we claim that this transition can be treated semi-classically once the fluctuation of the scalars fields are protected by the induced de Sitter (mass$)^2$. 
The structure of the Minkowski vacuum is generically different from that of the minimal Starobinsky model.
The classical Minkowski vacuum can be degenerate in certain cases so that the quantum mass terms are becoming relevant,
since they will be needed to determine the correct quantum vacuum of the theory. 

The third step of this work is devoted to the supersymmetric extension of the minimal $SO(1,1+n)$ model. This was achieved in section $3$ in the framework of $N=1$ supergravity where we show that the $SO(1,1+n)/SO(1+n)$ manifold is minimally extended to a   K\"ahlerian manifold ${\cal M}_1=SU(1,1+n)/ U(1)\times SU(1+n)$.  
The observation here is that ${\cal M}_1$ is nothing but the
no-scale manifold which controls the kinetic part of the scalars and absorbs in a ``magic" way the negative contribution of the supergravity auxiliary fields, namely the well known $-3m^2_{3/2}$ term of the supergravity potential. This ``magic'' property of ${\cal M}_1 $ leads to the positivity properties the $F$ part of the potential {\it independently} if supersymmetry is broken or not, provided that the no-scale modulus does not appear in the superpotential. As a result, the supergravity $F$ part of the potential looks like the global supersymmetric potential dressed by the no-scale modulus $\phi$ in a multiplicative sense. This positivity property is explained in section $3$ and is further extended in more general no-scale manifolds coming from strings in section 5. In fact, this is the main reason for the failure of the majority of the models proposed in the literature when trying to find metastable de Sitter vacua by modifying in an ad-hoc way the $T$-dependent superpotential. We show by presenting explicit examples, both 
in supergravity in section 3 but also in superstring induced supergravity models in section 5,
 that the only natural way to resolve the instabilities of a $T$-dependent superpotential can be achieved when a non trivial Fayet-Iliopoulos term is created by a gauging of a $U(1)_R$ symmetry. 
 %We show that at the classical level this is possible by presenting explicit representative examples both in supergravity in section 3 but also in superstring induced supergravity models in section 5.  
 However, this is still not yet an achievement, since all possible $U(1)_R$ gaugings, able to create Fayet-Iliopoulos terms, are anomalous! This means that at the quantum level the theory is inconsistent. It implies that the gauging of $U(1)_R$ alone is not enough and has to be extended in a consistent way by gauging simultaneously some of the isometries of the scalar manifold. This leads to two generic extensions of $U(1)_R$, namely to the $U(1)_d$ and $U(1)_t$ gaugings, involving the dilatation symmetry or the axionic symmetry of the scalar manifold.  Both of them cure the anomaly by screening the $D$-term contribution in a scale invariant manner 
or, respectively, in a non-scale invariant manner. In both cases the de Sitter vacuum is destabilized either in a rapid or slow way 
giving in both cases inflationary behavior during the transition to the flat Minkowski vacuum.  
%In the dilatation case (where the classical scale symmetry is preserved) the resulting off-shell potential of the theory turns out to be  quadratic in terms of $B$ (the imaginary component of the no-scale modulus $T$),
%while, in the $U(1)_t$ case the dressing of the $D$-terms appears  the no-scale modulus $\phi$ in a non scale invariant manner as have been explicitly shown in sections 4 and 5.  

Concerning the applications in cosmology, for the case of the $U(1)_d$ resolution of the anomaly, the de Sitter instability is relatively fast and is given in terms of a polynomial potential of degree two in terms of $B$ and degree four in terms of $\Phi$. This kind of inflationary potentials has to be treated as two component chaotic inflationary model.  Generically in this
 kind of inflationary potentials the predictions for the ratio $r$ of the tensor to scalar fluctuations is relatively large $r$=0.2 to 0.01. These models will be relevant if the BICEP2 preliminary results turn out to be confirmed.  

In section 5 we investigated the superstring vacuum structure in the generic case where three anomalous 
$U(1)^A_R$ are involved. The resolution of the anomaly via  $U(1)^A_t$  induces a slow transition from de Sitter to flat Minkowski space via three, two or one
 inflaton components. We postulate that in superstring theory the resolution of the various anomalous $U(1)^A_R$ is generic in  all string compactifications.  

Namely we show that $N=1$ superstring constructions provide metastable de Sitter vacua supported by Fayet-Iliopoulos $D_R$-terms associated to the several anomalous $U(1)^A_R$  gauge symmetries. 
The superstring resolution of these anomalies is achieved via 
local axion shifts, which promote the several $U(1)_R$ symmetries to $ U(1)_{t}$ or $U(1)_d$ symmetries.
  
 \section{Acknowledgement*}
We are grateful to S. Ferrara for fruitful discussions, especially for some clarifications concerning the Fayet-Iliopoulos terms in supergravity and also about some of 
the restrictions concerning the equivalence of the  ``minimal'' versus ``non-minimal'' formulation of $N=1$ supergravity. 
We also appreciated fruitful discussions with V. Mukhanov, H. Partouche, K. Stelle, J. Troost and  B.~de~ Wit.
 C. K. acknowledges the theoretical physics group of Ludwig Maximilians University and 
Max-Planck-Institute in Munich for hospitality and he
likes to thank the University of Cyprus for hospitality, where part of the work was done.
C.K. and D.L. like to thank the theory department of CERN, where part of the work was done.
N.T.  acknowledge the Laboratoire de Physique Th\'eorique of Ecole Normale Sup\'erieure for hospitality. 
The work of C.K. and N.T. is also supported by the CEFIPRA/IFCPAR 4104-2 project and a PICS France/Cyprus. 
This research is also supported by the Munich Excellence Cluster for Fundamental Physics ``Origin and the Structure of the Universe" and by the
ERC Advanced Grant ``Strings and Gravity" (Grant No. 32004).
The work of C.K. is partially supported the Gay Lussac-Humboldt Research Award 2014, at the Ludwig Maximilians University and Max-Planck-Institute for Physics. 

%%%%%%%%%%%%%%%%%%%%%%%%%%%%%%%%%%%%%
%%%%%%%%%%%%%%%%%%%%%%%%%%%%
%%%%%%%%%%%%%%%%%%%%%%%%%%%%%%%%%%%%
%%%%%%%%%%%%%%%%%%%%%%%%%%%%%%%%%%%%

\end{document}